\documentclass[fleqn,3p,onecolumn,11pt,nofootinbib,showkeys,showpacs]{revtex4-1}

\usepackage[small,justification=centerlast]{caption}
\usepackage{graphicx}
\usepackage{bm}
\usepackage{mathrsfs}
\usepackage[latin1]{inputenc}
\usepackage{stmaryrd}
\usepackage{array}
\usepackage{psfrag}
\usepackage{dsfont}
\usepackage{epsfig}
\usepackage{titletoc}
\usepackage{float}   
\usepackage{wrapfig} 
\usepackage{amsmath}
\usepackage{amssymb,stmaryrd}
\usepackage{psfrag}
\usepackage{cancel}
\usepackage[dvips]{feynmp}
\usepackage{upgreek}
\usepackage{subfig}
\usepackage{tabularx}
\usepackage{xcolor}
\usepackage{units}

\usepackage{textfit} 

\usepackage{hyperref}

\renewcommand{\eqref}[1]{\mbox{Eq.~(\ref{#1})}}
\newcommand{\tabref}[1]{\mbox{Tab.~\ref{#1}}}
\newcommand{\figref}[1]{\mbox{Fig.~\ref{#1}}}
\newcommand{\secref}[1]{\mbox{Sec.~\ref{#1}}}

\newcommand{\appref}[1]{\mbox{App.~\ref{#1}}}

\begin{document}

\title{Quantum field theoretic properties of Lorentz-violating operators of nonrenormalizable dimension in the fermion sector}

\author{M. Schreck} \email{mschreck@indiana.edu}
\affiliation{Indiana University Center for Spacetime Symmetries, Indiana University, Bloomington, Indiana 47405-7105}

\begin{abstract}
In the current paper the properties of a quantum field theory based on certain sets of Lorentz-violating coefficients in the
nonminimal fermion sector of the Standard-Model Extension are analyzed. In particular, three families of coefficients are considered,
where two of them are {\em CPT}-even and the third is {\em CPT}-odd. As a first step the modified fermion dispersion relations
are obtained. Then the positive- and negative-energy solutions of the modified Dirac equation and the fermion propagator are derived.
These are used to demonstrate the validity of the optical theorem at tree-level, which provides a cross-check for the results obtained.
Furthermore unitarity is examined and seems to be valid for the first set of {\em CPT}-even coefficients. However for the remaining
sets certain issues with unitarity are found. The article demonstrates that the adapted quantum field theoretical methods at tree-level
work for the nonminimal, Lorentz-violating framework considered. Besides, the quantum field theory based on the first family of {\em CPT}-even
coefficients  is most likely well-behaved at lowest order perturbation theory. The results are important for future phenomenological
investigations carried out in the context of field theory, e.g., the computation of decay rates and cross sections at tree-level.
\end{abstract}

\maketitle

\newpage
\setcounter{equation}{0}
\setcounter{section}{0}
\renewcommand{\theequation}{\arabic{section}.\arabic{equation}}

\section{Introduction}
\setcounter{equation}{0}

Investigating Lorentz invariance violation has become more and more attractive in recent years. A possible violation of this
fundamental symmetry of the laws of nature is motivated by physics at the Planck scale such as string theory
\cite{Kostelecky:1988zi,Kostelecky:1991ak,Kostelecky:1994rn,oai:arXiv.org:hep-th/9605088},
loop quantum gravity \cite{Bojowald:2004bb,Gambini:1998it}, field theory on noncommutative spacetimes \cite{Carroll:2001ws}, spacetime
foam models \cite{Klinkhamer:2003ec,Bernadotte:2006ya} and quantum field theory on spacetimes with a nontrivial topology
\cite{Klinkhamer:1998fa,Klinkhamer:1999zh}. An effective description of Lorentz symmetry violation for
energies that are much smaller than the Planck scale is provided by the Standard-Model Extension (SME) \cite{Colladay:1998fq}. The latter
forms a test framework for experimental searches \cite{Kostelecky:2008ts} for Lorentz symmetry violation and it allows allows one to investigate
the properties of
quantum field theories based on certain sectors of this framework. The SME includes all operators of Standard Model fields that are invariant
with respect to the gauge group $\mathit{SU}(3)_c\times \mathit{SU}(2)_L\times \mathit{U}(1)_Y$ and violate particle Lorentz invariance.
The minimal version is restricted to only power-counting renormalizable operators, whereas the nonminimal version also contains operators of
higher dimension \cite{Kostelecky:2009zp,Kostelecky:2011gq,Kostelecky:2013rta}.

In a series of articles quantum field theories based on a Lorentz-violating modification of the minimal photon sector were examined
\cite{oai:arXiv.org:hep-ph/0101087,Casana-etal2009,Casana-etal2010,Klinkhamer:2010zs,Schreck:2011ai,Schreck:2013gma,Colladay:2014dua}.
They applied to characteristics of the modified photon propagator, the polarization vectors, unitarity, and microcausality. In a recent paper
these methods are even employed for a quantum field theory based on a special set of operators of the nonminimal photon sector  \cite{Schreck:2013kja}.
Although there is more knowledge to be gained for the nonminimal photon sector, some of the main properties of a quantum field theory based on
an isotropic operator of nonrenormalizable dimension were obtained in the latter reference. To extend the picture the current paper is devoted
to similar investigations for the nonminimal fermion sector.
Before delving into phenomenological calculations, the properties of the underlying quantum field theory should be investigated,
which is one of the goals of the paper. Furthermore the results obtained such as the modified spinors and propagators are ready to be used in
phenomenology as long as the quantum theory proves to be consistent.
Note that the majority of both experimental and theoretical investigations performed to date has been restricted to the minimal fermion sector of
the SME.

The basis of a quantum field theory of spin-1/2 particles is formed by the Dirac equation. Dirac introduced the equation that is named after
him in 1928 for several reasons \cite{Dirac:1928hu}. First, the number of stationary states in hydrogen-like atoms were observed to be twice
what the quantum theory of a pointlike electron without internal quantum numbers would suggest. To account for this doubling of states
quantum-mechanical spin was introduced by Pauli and Darwin (see \cite{Dirac:1928hu} and references therein). However it was unsatisfactory
that the spin had to be introduced by hand and did not arise naturally from the theory. Second, the relativistic wave equation proposed by Klein
and Gordon evidently would describe electrons of both negative and positive charge where the latter are associated with a negative energy.
Classically these solutions could be discarded but quantum mechanically transitions between states with negative and positive charge
could be induced by perturbation, which is not observed in nature. These, amongst other problems, were solved by the Dirac equation, which
incorporates special relativity into quantum mechanics and, therefore, naturally describes the electron spin. Beyond the context of quantum
mechanics it was reinterpreted and used in quantum field theory to describe fermions with spin 1/2.

A modified version of the Dirac Lagrange density leading to a modified Dirac equation forms the foundation of the Lorentz-violating fermion
sector. Quantum field theoretic properties of the minimal fermion sector such as microcausality and stability were investigated in
\cite{Kostelecky:2000mm}. Furthermore the implications of a nonvanishing torsion coupling to the fermion sector were considered
in~\cite{Kostelecky:2007kx} where the occuring terms were stated, classified, and embedded into the minimal fermionic SME. Therefore already
existing bounds on minimal fermionic coefficients could be reinterpreted as bounds on the torsion tensor coefficients.

In the recent years a large collection of bounds have been obtained in the minimal SME fermion sector, especially by considering the gravitational
interaction. Such an approach is very reasonable, since certain coefficients of the minimal fermion sector are not observable in Minkowski spacetime,
even if they lie several orders of
magnitudes above the current experimental bounds of observable coefficients. In \cite{Kostelecky:2008in} it was shown that in the presence of
gravity the fluctuational part of a special sample of coefficients can be detected if they couple to the gravitational field. This was then
exploited to obtain several bounds on these coefficients from experiment.

The gravitational interaction of the Lorentz-violating minimal fermion sector was extensively studied in \cite{Kostelecky:2010ze}, where a large
number of bounds on the coefficients was determined. This list of bounds was extended in \cite{Tasson:2012nx} by considering an additional
experimental setup that had a priori not been designed for experiments in a gravitational background. Furthermore, in
\cite{Tasson:2010nr,Tasson:2012au} it is proposed that fermionic Lorentz-violating coefficients could be constrained by antimatter tests in
gravitational physics.

The current paper is organized as follows. Section~\ref{sec:introduction-theory} provides the action
of the nonminimal free SME fermion sector and restricts it to the operators that shall be investigated throughout the paper. In
\secref{sec:fermion-dispersion-laws} the modified fermion dispersion laws will be examined and \secref{sec:dirac-spinors} is dedicated to
the properties of the modified Dirac spinors. Section~\ref{sec:fermion-propagator} deals with the fermion propagator and the optical theorem
at tree-level, which relates the propagator to the sums of matrices in spinor space formed of the positive- and negative-energy
spinors, respectively. Finally, in \secref{sec:application-other-coefficients}
the analysis is extended to alternative sets of Lorentz-violating operators of the nonminimal fermion sector. The results are summarized and
discussed in \secref{sec:conclusion}. Calculational details are presented in the appendix. Throughout the article natural units with $\hbar=c=1$
will be used unless otherwise stated.

\section{Introduction of the theory}
\label{sec:introduction-theory}
\setcounter{equation}{0}

The theory considered is a Lorentz-violating extension of the free Standard-Model fermion sector~\cite{Kostelecky:2013rta}, which is based on the
following action:
\begin{equation}
\label{eq:action-fermion-sector}
S=\int_{\mathbb{R}^4} \mathrm{d}^4x\,\mathcal{L}\,,\quad \mathcal{L}=\frac{1}{2}\overline{\psi}\left(\gamma^{\mu}\mathrm{i}\partial_{\mu}-m_{\psi}\mathds{1}_4+\widehat{\mathcal{Q}}\right)\psi+\text{H.c.}\,,
\end{equation}
with the standard Dirac field $\psi$, the Dirac conjugate field $\overline{\psi}\equiv\psi^{\dagger}\gamma^0$, the fermion mass $m_{\psi}$, and the unit matrix $\mathds{1}_4$ in spinor space.
The standard gamma matrices $\gamma^{\mu}$ for $\mu=0 \dots 3$ satisfy the Clifford algebra $\{\gamma^{\mu},\gamma^{\nu}\}=2\eta^{\mu\nu}\mathds{1}_4$
with the Minkowski metric $(\eta^{\mu\nu})=\mathrm{diag}(1,-1,-1,-1)$. The part $\widehat{\mathcal{Q}}$ contains all Lorentz-violating operators to
arbitrary operator dimension that are compatible with the fermion sector.
The Lagrange density is written such that the corresponding Hamilton operator is Hermitian.

The modification $\widehat{\mathcal{Q}}$ represents an infinite sum of different composite operators, which are composed of momenta and
Lorentz-violating component coefficients. These operators can be grouped in different classes according
to their properties under (proper) observer Lorentz transformations and {\em C}, {\em P}, {\em T} transformations. This was done in Tab.~I of
\cite{Kostelecky:2013rta}. In what follows, the properties of different classes of operators shall be investigated, including $\widehat{m}$,
$\widehat{c}^{\,\mu}$, and $\widehat{f}$. The simplest one is undoubtedly $\widehat{m}$ in the first line of the table previously referred to.
Therefore, in the following section the action of \eqref{eq:action-fermion-sector} will be restricted to this particular operator.
In \secref{sec:application-other-coefficients} the methods chosen will be applied to $\widehat{c}^{\,\mu}$ and $\widehat{f}$, as well.

\subsection{Scalar operator}

The operator $\widehat{m}$ is {\em CPT}-even and does not have a dimension-4 field operator equivalent. Then $\widehat{\mathcal{Q}}$ is
given by
\begin{equation}
\label{eq:dimensionality-expansion-mhat}
\widehat{\mathcal{Q}}=-\widehat{m}\mathds{1}_4\,,\quad \widehat{m}=\sum_{\substack{d \text{ odd} \\ d\geq 5}}^{\infty} m^{(d)\alpha_1\dots \alpha_{(d-3)}}p_{\alpha_1}\dots p_{\alpha_{(d-3)}}\,.
\end{equation}
The $\alpha_1\dots \alpha_{(d-3)}$ are Lorentz indices in this context. Since $\widehat{m}$ does not have any free Lorentz
indices, it is a scalar under observer Lorentz transformations by construction. However the additional momentum dependence makes it transform
nontrivially under particle Lorentz transformations.
In \cite{Kostelecky:2013rta} these expansions are directly defined in momentum space. Then no additional signs have to be taken into account,
which simplifies the notation.
The number $d$ does not give the mass dimension of the component coefficients but the dimensionality of the corresponding field operator, which is
$\overline{\psi}\partial_{\alpha_1}\dots \partial_{(d-3)}\psi$ in configuration space.
In what follows, the dimensional expansion of $\widehat{m}$ in \eqref{eq:dimensionality-expansion-mhat} will be restricted to the dimension-5
field operator $\overline{\psi}\partial_{\alpha_1}\partial_{\alpha_2}\psi$. The corresponding 10 component coefficients $m^{(5)\alpha_1\alpha_2}$
have mass dimension $-1$, so we have to consider
\begin{equation}
\widehat{m}=\widehat{m}(p^0,\mathbf{p})=m^{(5)\alpha_1\alpha_2}p_{\alpha_1}p_{\alpha_2}\,.
\end{equation}
The arguments of $\widehat{m}$ will be suppressed unless caution is required.
The advantage of using the particular operator $\widehat{m}$ is that we can solely concentrate on effects that are characteristic for higher-dimensional
operators and that do not have an equivalent for relevant or marginal operators (operators with dimension smaller than 4 and those
with dimension equal to 4, respectively). It is reasonable to follow the same
steps as in \cite{Schreck:2013kja} and to consider three different sectors of the ten component coefficients:
\begin{equation}
m^{(5)}=\left(\begin{array}{c|ccc}
m^{(5)00} & m^{(5)01} & m^{(5)02} & m^{(5)03} \\
\hline
m^{(5)01} & m^{(5)11} & m^{(5)12} & m^{(5)13} \\
m^{(5)02} & m^{(5)12} & m^{(5)22} & m^{(5)23} \\
m^{(5)03} & m^{(5)13} & m^{(5)23} & m^{(5)33}
\end{array}\right)\,.
\end{equation}
The sector consisting of the single coefficient $m^{(5)00}$ will be called ``temporal,'' the sector made up of the three coefficients
$m^{(5)0i}$ for $i=1\dots 3$ will be named ``mixed,'' and the set of the remaining coefficients $m^{(5)ij}$ for $i$, $j=1\dots 3$ will be
denoted as ``spatial.''

\section{Modified fermion dispersion laws}
\label{sec:fermion-dispersion-laws}
\setcounter{equation}{0}

In the current section the modified fermion dispersion relations shall be computed and their properties will be discussed.
The left-hand side of equation (39) with the definition (35) in \cite{Kostelecky:2013rta} states the general off-shell dispersion
relation\footnote{In the literature the expression ``off-shell dispersion relation'' is sometimes used for the polynomial in $p^0$,
whose zeros give the dispersion relations for an on-shell particle.} for the Lorentz-violating fermion sector defined by the action of
\eqref{eq:action-fermion-sector}. For the special case considered here we have that $\widehat{\mathcal{S}}_{\pm}=-(m_{\psi}+\widehat{m})$,
$\widehat{\mathcal{V}}^{\mu}_{\pm}=p^{\mu}$, $\widehat{\mathcal{T}}^{\mu\nu}_{\pm}=0$, which leads to
\begin{equation}
\label{eq:off-shell-dispersion-law}
p^2-(m_{\psi}+\widehat{m})^2=0\,,
\end{equation}
where $p=(p^0,\mathbf{p})\equiv (\widetilde{E}_{\psi},\mathbf{p})$ is the fermion four-momentum with the spatial momentum $\mathbf{p}$. The solutions
of \eqref{eq:off-shell-dispersion-law} with respect to $\widetilde{E}_{\psi}$ correspond to the modified dispersion relations of a fermion.\footnote{
Tildes are used throughout the paper to distinguish modified quantities such as the particle energy from the standard results for these quantities.
Particles with modified properties will carry a tilde as well to emphasize that this particular type of particles is affected by Lorentz violation and to
oppose them to particles, which remain unaffected.}
There are both zeros $\widetilde{E}_{\psi}^{(>)}>0$ and $\widetilde{E}_{\psi}^{(<)}<0$ where only the positive-energy solutions will be given in what follows. For
the temporal sector they read:
\begin{subequations}
\begin{equation}
\label{eq:fermion-energy-temporal}
\widetilde{E}_{\psi;1,2}^{(\mathrm{temp})}=\frac{\sqrt{1-2m^{(5)00}m_{\psi}\mp\sqrt{1-4m^{(5)00}\left(m_{\psi}+m^{(5)00}\mathbf{p}^2\right)}}}{\sqrt{2}|m^{(5)00}|}\,, \\[2ex]
\end{equation}
with the expansions
\begin{align}
\label{eq:fermion-energy-temporal-perturbed}
\widetilde{E}_{\psi;1}^{(\mathrm{temp})}&=\sqrt{\mathbf{p}^2+m_{\psi}^2}(1+m^{(5)00}m_{\psi})+\mathcal{O}[(m^{(5)00})^2]\,, \\[2ex]
\label{eq:fermion-energy-temporal-spurious}
\widetilde{E}_{\psi;2}^{(\mathrm{temp})}&=\frac{1}{|m^{(5)00}|}-m_{\psi}\,\mathrm{sgn}(m^{(5)00})-\frac{1}{2}(\mathbf{p}^2+2m_{\psi}^2)|m^{(5)00}|+\mathcal{O}[(m^{(5)00})^2]\,,
\end{align}
where
\begin{equation}
\label{eq:sign-function}
\mathrm{sgn}(x)=\left\{\begin{array}{rcc}
1 & \text{for} & x>0\,, \\
0 & \text{for} & x=0\,, \\
-1 & \text{for} & x<0\,. \\
\end{array}
\right.
\end{equation}
\end{subequations}
Hence there are two modified dispersion laws. The first is a perturbation of the standard dispersion relation for a Dirac fermion with spatial
momentum $\mathbf{p}$ and mass $m_{\psi}$. However the second does evidently not have a limit for a vanishing Lorentz-violating coefficient
$m^{(5)00}$. Instead there is an energy gap, which is inversely proportional to the coefficient $m^{(5)00}$. The latter dispersion law may
become important for large fermion momenta indicating that it is related to Planck scale physics. Such
dispersion relations can be considered as spurious for momenta that are much smaller than the Planck scale. They also appear in the
context of the nonminimal photon sector (cf. \cite{Kostelecky:2009zp,Schreck:2013kja}) and how to deal with them will be described later.

For the mixed sector one obtains:
\begin{subequations}
\begin{align}
\label{eq:fermion-energy-mixed}
\widetilde{E}_{\psi}^{(\mathrm{mixed})}&=\frac{\mathbf{p}^2+m_{\psi}^2}{\sqrt{\left[1-(\widehat{m}_1)^2\right]\mathbf{p}^2+m_{\psi}^2}+\widehat{m}_1m_{\psi}}\,,\quad \widehat{m}_1=2m^{(5)0i}p^i\,, \\[2ex]
\label{eq:fermion-energy-mixed-expanded}
\widetilde{E}_{\psi}^{(\mathrm{mixed})}&=\sqrt{\mathbf{p}^2+m_{\psi}^2}-\widehat{m}_1m_{\psi}+\mathcal{O}[(\widehat{m}_1)^2]\,.
\end{align}
\end{subequations}
Here no spurious dispersion law appears in contrast to the mixed sector of the particular set of nonminimal photon coefficients considered
in \cite{Schreck:2013kja}. Last but not least, for the spatial sector the modified dispersion law is given by:
\begin{subequations}
\begin{align}
\label{eq:fermion-energy-spatial}
\widetilde{E}_{\psi}^{(\mathrm{spatial})}&=\sqrt{\mathbf{p}^2+(m_{\psi}+\widehat{m}_2)^2}\,,\quad \widehat{m}_2=m^{(5)ij}p^ip^j\,, \\[2ex]
\label{eq:fermion-energy-spatial-expanded}
\widetilde{E}_{\psi}^{(\mathrm{spatial})}&=\sqrt{\mathbf{p}^2+m_{\psi}^2}\left(1+\frac{m_{\psi}}{\mathbf{p}^2+m_{\psi}^2}\widehat{m}_2\right)+\mathcal{O}[(\widehat{m}_2)^2]\,.
\end{align}
\end{subequations}
Also for the spatial sector there is no spurious dispersion relation in accordance to the nonminimal photon theory discussed in the latter
reference.

For fermions the negative-energy solutions have a physical meaning as well. They will not be given explicitly but they are related to
the positive-energy solutions as follows: $\widetilde{E}_{\psi}^{(>)}(\mathbf{p},m^{(5)\alpha_1\alpha_2})=-\widetilde{E}_{\psi}^{(<)}(-\mathbf{p},m^{(5)\alpha_1\alpha_2})$.
Now the negative-energy solutions have to be reinterpreted. The basis for this is the idea of the Dirac sea telling us that the vacuum
consists of an infinite number of filled negative-energy states. The corresponding positive-energy excitations, which are understood as positively
charged holes in the Dirac sea, are interpreted as antiparticles.
According to the Feynman-St\"{u}ckelberg interpretation a negative-energy particle propagating backwards in time, i.e., having four-momentum
$(p^{\mu})=(-p^0,-\mathbf{p})^T$, is interpreted as a positive-energy antiparticle propagating forwards in time with
$(p^{\mu})=(p^0,\mathbf{p})^T$. This concept, which describes the behavior of antiparticles in the framework of the
Dirac sea, is very helpful to understand how the positive energy of the physical antiparticle can be obtained from the negative-energy solution
of the Dirac equation. This is possible by reinterpreting
$p^0=\widetilde{E}_{\psi}^{(<)}(\mathbf{p},m^{(5)\alpha_1\alpha_2})$ with $p^{\mu}\mapsto -p^{\mu}$ where the latter transforms to
$p^0=\widetilde{E}_{\psi}^{(>)}(\mathbf{p},m^{(5)\alpha_1\alpha_2})$. While
$\widetilde{E}_{\psi}^{(>)}(\mathbf{p},m^{(5)\alpha_1\alpha_2})$ is the energy of a spin-1/2 matter particle, this can also be understood as
the energy of the corresponding antimatter particle.
Since the operator $\widehat{m}$, which is closely linked to the fermion mass $m_{\psi}$, is {\em CPT}-even \cite{Kostelecky:2013rta},
the sign of the corresponding component coefficients is not reversed when considering the negative-energy solutions. As a result, the particle and antiparticle energies are equal. This is
in accordance to the corresponding rules for the minimal fermion sector \cite{Kostelecky:2000mm}.

It can be checked that the expansions of Eqs.~(\ref{eq:fermion-energy-temporal-perturbed}), (\ref{eq:fermion-energy-mixed-expanded}), and
(\ref{eq:fermion-energy-spatial-expanded}) at first order in Lorentz violation are in agreement with the upper $2\times 2$ block of Eq.~(59)
in \cite{Kostelecky:2013rta} for particles and the reinterpreted lower $2\times 2$ block for antiparticles.

\section{Modified Dirac spinors}
\label{sec:dirac-spinors}
\setcounter{equation}{0}

The Lagrange density in \eqref{eq:action-fermion-sector} leads to a modified Dirac equation for the spinor field $\psi$ that is given as follows:
\begin{equation}
\label{eq:modified-dirac-equation}
(\cancel{p}-m_{\psi}\mathds{1}_4+\widehat{\mathcal{Q}})\psi=0\,,\quad (\gamma^{\mu})=(\gamma^0,\boldsymbol{\gamma})^T\,,\quad \boldsymbol{\gamma}=(\gamma^1,\gamma^2,\gamma^3)^T\,.
\end{equation}
After having obtained the modified fermion dispersion laws in the last section the solutions of the modified Dirac equation will be
determined in the current section. The procedure described in \cite{Kostelecky:2013rta} shall be used for this purpose.
First of all, according to the latter reference we choose a special representation of gamma-matrices --- the chiral representation, in
which the $\gamma^{0,1,2,3}$ are block off diagonal and $\gamma^5$ is diagonal. Explicitly the matrices are given by:
\begin{subequations}
\begin{equation}
\gamma^0=\begin{pmatrix}
0 & \mathds{1}_2 \\
\mathds{1}_2 & 0 \\
\end{pmatrix}\,,\quad \gamma^{1,2,3}=\begin{pmatrix}
0 & \sigma^{1,2,3} \\
-\sigma^{1,2,3} & 0 \\
\end{pmatrix}\,,\quad \gamma^5=\mathrm{i}\gamma^0\gamma^1\gamma^2\gamma^3=\begin{pmatrix}
-\mathds{1}_2 & 0 \\
0 & \mathds{1}_2 \\
\end{pmatrix}\,,
\end{equation}
with the Pauli matrices
\begin{equation}
\label{eq:pauli-matrices}
\sigma^1=\begin{pmatrix}
0 & 1 \\
1 & 0 \\
\end{pmatrix}\,,\quad \sigma^2=\begin{pmatrix}
0 & -\mathrm{i} \\
\mathrm{i} & 0 \\
\end{pmatrix}\,,\quad \sigma^3=\begin{pmatrix}
1 & 0 \\
0 & -1 \\
\end{pmatrix}\,,
\end{equation}
\end{subequations}
and the two-dimensional unit matrix $\mathds{1}_2$. For the particular case considered the solutions of the Dirac equation can be
determined from Eqs. (51), (53), and (54) in \cite{Kostelecky:2013rta}. The procedure will be briefly reviewed for the standard
Dirac equation first, i.e., \eqref{eq:modified-dirac-equation} with $\widehat{\mathcal{Q}}=0$. The initial step is to construct a unitary transformation
matrix $U$ depending on an energy scale $E\geq 0$ and a mass scale $m$, which looks as follows:
\begin{equation}
\label{eq:diagonalization-matrix}
U(E,m,\mathbf{p})=V\cdot W(E,m,\mathbf{p})\,,\quad V=\frac{\mathds{1}_4+\gamma^0\gamma^5}{\sqrt{2}}\,,\quad W(E,m,\mathbf{p})=\frac{(E+m)\mathds{1}_4+\mathbf{p}\cdot\boldsymbol{\gamma}}{\sqrt{2E(E+m)}}\,.
\end{equation}
For the standard theory with zero Lorentz violation, $m$ corresponds to the fermion mass $m_{\psi}$ and $E$ to the fermion energy
$E_{\psi}=\sqrt{\mathbf{p}^2+m_{\psi}^2}$. Using the matrix $U$, the Dirac operator can be diagonalized leading to the following
eigenvalue problem for the Hamiltonian $H$:
\begin{equation}
\label{eq:energy-eigenvalue-problem}
(E\mathds{1}_4-H)U\psi=0\,,\quad H=-\gamma^5E_{\psi}=\begin{pmatrix}
E_{\psi}\mathds{1}_2 & 0 \\
0 & -E_{\psi}\mathds{1}_2 \\
\end{pmatrix}\,.
\end{equation}
Once the Dirac matrix has been brought to this form it is straightforward to obtain its solutions for the transformed spinor $\psi'\equiv U\psi$.
However since the interest lies in the solutions for the spinor $\psi$, the obtained solutions have to be transformed back with $U^{\dagger}$. Note that this
procedure looks very similar to a Foldy-Weythousen transformation, cf. Eq. (17) in \cite{Foldy:1949wa} (with
$\beta\boldsymbol{\alpha}=\boldsymbol{\gamma}$ and $\beta^2=1$). The difference is that Foldy and Weythousen considered different transformation
matrices for positive- and negative-energy solutions of the Dirac equation, which are given by their Eqs. (17) and (18). On the contrary, in
\cite{Kostelecky:2013rta} only the single transformation matrix of \eqref{eq:diagonalization-matrix} is considered, since the negative-energy solutions
are obtained from the positive-energy ones by a reinterpretation.

According to \cite{Kostelecky:2013rta}, the diagonalization of the Dirac operator with the transformation given by
\eqref{eq:diagonalization-matrix} still works in case of a nonvanishing operator $\widehat{m}$. The only modification is that
$E_{\psi}$ has to be resplaced by $\widetilde{E}_{\psi}$ and $m_{\psi}$ by $\widetilde{m}_{\psi}$ (defined below) in \eqref{eq:diagonalization-matrix}.
With the diagonalization performed, both the positive- and the negative-energy spinors of the modified Dirac equation can be
obtained. The positive-energy spinors $u$ are then given by:
\begin{subequations}
\label{eq:positive-energy-spinors}
\begin{align}
u^{(\alpha)}(\widetilde{E}_{\psi}^{(>)},\mathbf{p})&=\frac{1}{\sqrt{N_u^{(\alpha)}}}U^{\dagger}(\widetilde{E}_{\psi}^{(>)},\widetilde{m}_{\psi},\mathbf{p})u^{(\alpha)}(\widetilde{m}_{\psi},\mathbf{0})\,, \displaybreak[0] \\[2ex]
\label{eq:positive-energy-spinors-at-rest}
u^{(1)}(\widetilde{m}_{\psi},\mathbf{0})&=\begin{pmatrix}
\phi^{(1)} \\
\mathbf{0} \\
\end{pmatrix}\,,\quad u^{(2)}(\widetilde{m}_{\psi},\mathbf{0})=\begin{pmatrix}
\phi^{(2)} \\
\mathbf{0} \\
\end{pmatrix}\,,\quad \phi^{(1)}=\begin{pmatrix}
1 \\
0 \\
\end{pmatrix}\,,\quad \phi^{(2)}=\begin{pmatrix}
0 \\
1 \\
\end{pmatrix}\,, \displaybreak[0]\\[2ex]
\widetilde{m}_{\psi}&\equiv m_{\psi}+\widehat{m}(\widetilde{E}_{\psi}^{(>)},\mathbf{p})\,.
\end{align}
\end{subequations}
where $\alpha=1$, 2 and $\widetilde{E}_{\psi}^{(>)}$ is the positive fermion energy that is modified due to Lorentz violation (cf. Eqs. (\ref{eq:fermion-energy-temporal}),
(\ref{eq:fermion-energy-mixed}), and (\ref{eq:fermion-energy-spatial}), respectively, for the three different sectors considered). The spinors are a solution
of the modified Dirac equation: $(\cancel{p}-\widetilde{m}_{\psi}\mathds{1}_4)u^{(\alpha)}(p)=0$ with $p^0=\widetilde{E}_{\psi}^{(>)}$. The mormalization
$N_u^{(\alpha)}$ of the spinors is chosen such that
\begin{subequations}
\begin{align}
\overline{u}^{(\alpha)}(p)u^{(\beta)}(p)&=u^{(\alpha)\,\dagger}(p)\gamma^0u^{(\beta)}(p)=2\widetilde{m}_{\psi}\delta^{\alpha\beta}\,, \\[2ex]
u^{(\alpha)\,\dagger}(p)u^{(\beta)}(p)&=2\widetilde{E}_{\psi}^{(>)}\delta^{\alpha\beta}\,.
\end{align}
\end{subequations}
In what follows, the matrices formed from the spinors, e.g., $u^{(\alpha)}(p)\overline{u}^{(\alpha)}(p)$ will be denoted as ``spinor matrices.'' The sum over the
positive-energy spinor matrices then reads
\begin{equation}
\label{eq:completeness-relation-positive}
\sum_{\alpha=1,2} u^{(\alpha)}(p)\overline{u}^{(\alpha)}(p)=\cancel{p}+\widetilde{m}_{\psi}\mathds{1}_4\,.
\end{equation}
On the other hand, the negative-energy spinors are given by
\begin{subequations}
\label{eq:negative-energy-spinors}
\begin{align}
v^{(\alpha)}(\widetilde{E}^{(>)}_{\psi},\mathbf{p})&=\frac{1}{\sqrt{N_v^{(\alpha)}}}U^{\dagger}(\widetilde{E}^{(>)}_{\psi},\widetilde{m}_{\psi},-\mathbf{p})v^{(\alpha)}(\widetilde{m}_{\psi},\mathbf{0})\,, \\[2ex]
\label{eq:negative-energy-spinors-at-rest}
v^{(1)}(\widetilde{m}_{\psi},\mathbf{0})&=\begin{pmatrix}
\mathbf{0} \\
\chi^{(1)} \\
\end{pmatrix}\,,\quad v^{(2)}(\widetilde{m}_{\psi},\mathbf{0})=\begin{pmatrix}
\mathbf{0} \\
\chi^{(2)} \\
\end{pmatrix}\,,\quad \chi^{(1)}=\begin{pmatrix}
1 \\
0 \\
\end{pmatrix}\,,\quad \chi^{(2)}=\begin{pmatrix}
0 \\
1 \\
\end{pmatrix}\,, \\[2ex]
\widetilde{m}_{\psi}&=m_{\psi}+\widehat{m}(-\widetilde{E}^{(>)}_{\psi},-\mathbf{p})=m_{\psi}+\widehat{m}(\widetilde{E}^{(>)}_{\psi},\mathbf{p})\,.
\end{align}
\end{subequations}
Note the minus signs associated to the four-momentum components $p^0$ and $\mathbf{p}$. These spinors are a solution of the
modified Dirac equation $(\cancel{p}-\widetilde{m}_{\psi}\mathds{1}_4)v^{(\alpha)}(p)=0$ with $(p^{\mu})=(-\widetilde{E}_{\psi}^{(>)},-\mathbf{p})^T$.
Here the normalization $N_v^{(\alpha)}$ is chosen so that the following relationships hold:
\begin{subequations}
\begin{align}
\overline{v}^{(\alpha)}(p)v^{(\beta)}(p)&=v^{(\alpha)\,\dagger}(p)\gamma^0v^{(\beta)}(p)=-2\widetilde{m}_{\psi}\delta^{\alpha\beta} \\[2ex]
v^{(\alpha)\,\dagger}(p)v^{(\beta)}(p)&=-2\widetilde{E}_{\psi}\delta^{\alpha\beta}\,.
\end{align}
\end{subequations}
The sum over the negative-energy spinor matrices is given by
\begin{equation}
\label{eq:spinor-completeness-negative}
\sum_{\alpha=1,2} v^{(\alpha)}(p)\overline{v}^{(\alpha)}(p)=\cancel{p}-\widetilde{m}_{\psi} \mathds{1}_4\,.
\end{equation}
On the right-hand sides of Eqs.~(\ref{eq:completeness-relation-positive}) and (\ref{eq:spinor-completeness-negative}),
$(p^{\mu})=(\widetilde{E}_{\psi}^{(>)},\mathbf{p})^T$ is understood. An explicit derivation of all these relations can be found in
\appref{sec:spinors-completeness-explicit-mmu}.

\section{Modified fermion propagator and the optical theorem}
\label{sec:fermion-propagator}
\setcounter{equation}{0}

\begin{figure}[t]
\includegraphics{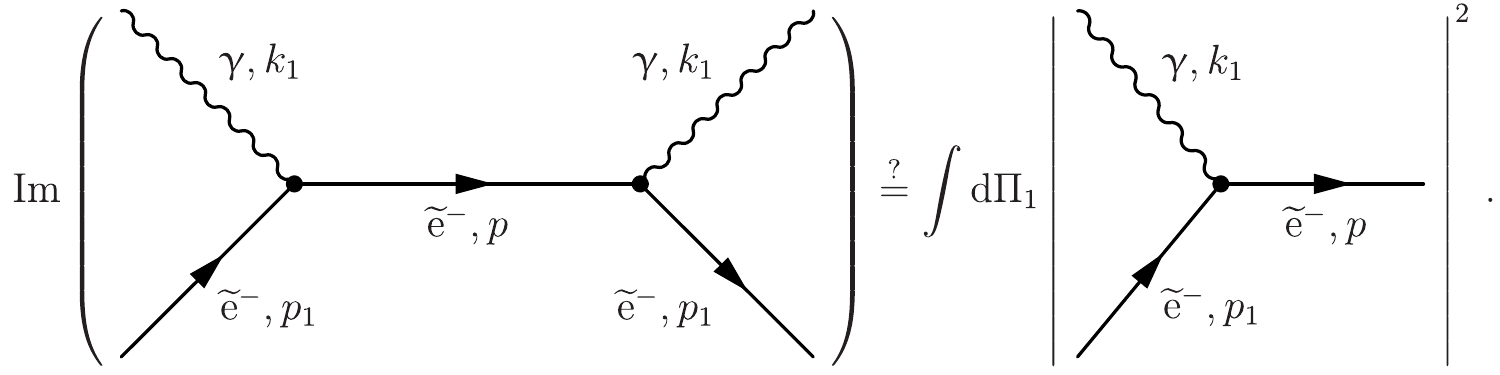}
\caption{Forward scattering amplitude of tree-level electron photon scattering that is related to the total cross section of electron
photon scattering, if the optical theorem is valid. A modified electron is denoted by $\widetilde{\mathrm{e}}^-$ and a photon by
$\upgamma$. The momenta are stated next to the particle symbols and the one-particle phase space is called $\mathrm{d}\Pi_1$.}
\label{fig:optical-theorem}
\end{figure}

Having obtained the modified spinors and sums over the spinor matrices in the last section, the fermion propagator will be computed in what follows.
The fermion propagator $S(p)$ is the inverse (modulo a factor of $\mathrm{i}$) of the operator $S^{-1}(p)\equiv \cancel{p}-m_{\psi}\mathds{1}_4+\widehat{\mathcal{Q}}$ that
appears in the modified Dirac equation: $S(p)S^{-1}(p)=S^{-1}(p)S(p)=\mathrm{i}\mathds{1}_4$. In the case of a nonvanishing operator $\widehat{m}$
it holds that $S^{-1}(p)=\cancel{p}-(m_{\psi}+\widehat{m})\mathds{1}_4$.
From the latter equation the propagator can be determined and it is expressed in terms of gamma matrices as follows:
\begin{equation}
\label{eq:electron-propagator}
S(p)=\frac{\mathrm{i}}{p^2-(m_{\psi}+\widehat{m})^2}\left[\cancel{p}+(m_{\psi}+\widehat{m})\mathds{1}_4\right]\,.
\end{equation}
As a good cross check for the electron propagator of \eqref{eq:electron-propagator} and the sum over the spinor matrices of \eqref{eq:completeness-relation-positive}
the optical theorem can be used. Therefore we consider a modified electron $\widetilde{\mathrm{e}}^-$ scattering at a standard
photon $\upgamma$ (Compton scattering). The forward scattering amplitude at tree-level corresponds to the left-hand side of the
equation shown in \figref{fig:optical-theorem} and it is denoted by $\mathcal{M}\equiv \mathcal{M}(\widetilde{\mathrm{e}}^-\upgamma \rightarrow \mathrm{\widetilde{e}^-}\upgamma)$.
If the optical theorem is valid, the forward scattering amplitude will be related to the total cross section of the process
$\mathrm{\widetilde{e}^-}\upgamma \rightarrow \mathrm{\widetilde{e}^-}$ at tree-level where a summation over the spins of the
final electron has to be performed.\footnote{At tree-level Compton scattering has an additional contribution with the two vertices
interchanged. The sum of both amplitudes is gauge invariant where a single contribution is not. Nevertheless to check the optical
theorem we restrict ourselves to only the first contribution. If the optical theorem is valid, the imaginary part of the first
amplitude will be related to the cross section of a physical process, which must be a gauge-invariant quantity. Hence the imaginary
part of the corresponding forward scattering amplitude is then gauge-invariant, as well.} If the spin state of the initial electron
is denoted as $\alpha$ and the polarization state of the initial photon as $\lambda$, the forward scattering amplitude reads:
\begin{equation}
\label{eq:forward-scattering-amplitude}
\mathcal{M}=-\int \frac{\mathrm{d}^4p}{(2\pi)^4}\,\delta^{(4)}(k_1+p_1-p)e^2\overline{u}^{(\alpha)}(p_1)\gamma^{\nu}\frac{\cancel{p}+\widetilde{m}_{\psi}\mathds{1}_4}{p^2-\widetilde{m}_{\psi}^2+\mathrm{i}\epsilon}\gamma^{\mu}u^{(\alpha)}(p_1)\varepsilon^{(\lambda)}_{\mu}(k_1)\overline{\varepsilon}^{(\lambda)}_{\nu}(k_1)\,.
\end{equation}
Here $u^{(\alpha)}(p_1)$ is a positive-energy spinor describing a modified electron in the spin state $\alpha$ and with four-momentum
$p_1$. These were obtained in the previous section and are given by \eqref{eq:positive-energy-spinors}. The elementary charge is $e$
and the Feynman propagator poles are treated with the usual $\mathrm{i}\epsilon$-prescription. The polarization vector of a standard
photon with polarization state $\lambda$ and momentum $k_1$ is named $\varepsilon_{\mu}^{(\lambda)}(k_1)$. Total energy-momentum
conservation of the process is encoded in the four-dimensional $\delta$-function.

The interest lies in the imaginary part of \eqref{eq:forward-scattering-amplitude}. First of all only the mixed and the spatial sector
of the theory, whose dispersion relations were obtained in \secref{sec:fermion-dispersion-laws}, are considered. These sectors are
characterized by a single positive and a negative fermion energy and the denominator of the corresponding propagator can be factorized
with respect to the poles as follows:
\begin{equation}
\label{eq:propagator-poles}
\frac{1}{p^2-\widetilde{m}_{\psi}^2+\mathrm{i}\epsilon}=\frac{1}{\left(p^0-\widetilde{E}_{\psi}^{(>)}+\mathrm{i}\epsilon\right)\left(p^0-\widetilde{E}_{\psi}^{(<)}-\mathrm{i}\epsilon\right)}\,,
\end{equation}
with the positive fermion energy $\widetilde{E}_{\psi}^{(>)}$ and the negative-energy counterpart $\widetilde{E}_{\psi}^{(<)}$. Due to
energy-momentum conservation only the pole with a positive real part, i.e., $p^0=\widetilde{E}_{\psi}^{(>)}-\mathrm{i}\epsilon$
contributes to the imaginary part. Interpreting the propagator as a distribution, one can use the following relation to treat the
contributing pole where this procedure corresponds to applying Cutkosky's cutting rules \cite{Cutkosky:1960sp}:
\begin{equation}
\frac{1}{p^0-\widetilde{E}_{\psi}^{(>)}+\mathrm{i}\epsilon}=\mathcal{P}\frac{1}{p^0-\widetilde{E}_{\psi}^{(>)}}-\mathrm{i}\pi\delta\left(p^0-\widetilde{E}_{\psi}^{(>)}\right)\,.
\end{equation}
Here the first term involves the principal value $\mathcal{P}$, which is purely real. The second summand is imaginary and forces $p^0$
to be equal to the fermion energy $\widetilde{E}_{\psi}^{(>)}$ in the integrand. With this input the imaginary part of \eqref{eq:forward-scattering-amplitude}
can be computed where additionally \eqref{eq:completeness-relation-positive} is used:
\begin{align}
2\mathrm{Im}(\mathcal{M})&=\int \frac{\mathrm{d}^3p}{(2\pi)^32\widetilde{E}_{\psi}^{(>)}}\delta^{(4)}(k_1+p_1-p)e^2\overline{u}^{(\alpha)}(p_1)\gamma^{\nu}(\cancel{p}+\widetilde{m}_{\psi}\mathds{1}_4)\gamma^{\mu}u^{(\alpha)}(p_1)\varepsilon_{\mu}^{(\lambda)}(k_1)\overline{\varepsilon}_{\nu}^{(\lambda)}(k_1) \notag \displaybreak[0]\\
&=\int \frac{\mathrm{d}^3p}{(2\pi)^32\widetilde{E}_{\psi}^{(>)}}\delta^{(4)}(k_1+p_1-p) \notag \displaybreak[0]\\
&\phantom{{}={}\int \frac{\mathrm{d}^3p}{(2\pi)^32\widetilde{E}_{\psi}^{(>)}}}\times e^2\overline{u}^{(\alpha)}(p_1)\gamma^{\nu}\Big[\sum_{\beta=1,2} u^{(\beta)}(p)\overline{u}^{(\beta)}(p)\Big]\gamma^{\mu} u^{(\alpha)}(p_1)\varepsilon_{\mu}^{(\lambda)}(k_1)\overline{\varepsilon}_{\nu}^{(\lambda)}(k_1) \notag \displaybreak[0]\\
&=\int \frac{\mathrm{d}^3p}{(2\pi)^32\widetilde{E}_{\psi}^{(>)}}\delta^{(4)}(k_1+p_1-p) \notag \\
&\phantom{{}={}\int \frac{\mathrm{d}^3p}{(2\pi)^32E_{\psi}^{(>)}}}\times \sum_{\beta=1,2} \left(\mathrm{i}e\overline{u}^{(\beta)}(p)\gamma^{\nu}u^{(\alpha)}(p_1)\varepsilon_{\nu}^{(\lambda)}(k_1)\right)^{\dagger}\mathrm{i}e\overline{u}^{(\beta)}(p)\gamma^{\mu}u^{(\alpha)}(p_1)\varepsilon^{(\lambda)}_{\mu}(k_1) \notag \displaybreak[0]\\
&=\int \frac{\mathrm{d}^3p}{(2\pi)^32\widetilde{E}_{\psi}^{(>)}}\delta^{(4)}(k_1+p_1-p) \sum_{\beta=1,2} |\widehat{\mathcal{M}}|^2.
\end{align}
Hence the imaginary part can be expressed with the matrix element $\widehat{\mathcal{M}}\equiv \mathcal{M}(\mathrm{\widetilde{e}^-}\upgamma \rightarrow \mathrm{\widetilde{e}^-})$
of the process on the right-hand side of the equation in \figref{fig:optical-theorem}. This shows that the optical theorem is valid for
this particular process. Note that this proof is rather general and no relations were used that are supposedly only valid for the process
considered.

An analogous computation can be done for spin-1/2 antimatter by considering the process $\widetilde{\mathrm{e}}^+\upgamma\rightarrow \widetilde{\mathrm{e}}^+\upgamma$,
with a modified positron $\widetilde{\mathrm{e}}^+$. Then the electron lines in the diagrams shown in \figref{fig:optical-theorem} have
to be replaced by positron lines. Since the momentum of the internal line flows in the opposite direction of the arrow on this line,
the propagator momentum is now $-p^{\mu}$ instead of $p^{\mu}$~\cite{Mandl:1986}. Furthermore a global factor of $-1$ has to be
considered due to the interchange of fermionic operators when applying Wick's theorem:
\begin{equation}
\label{eq:forward-scattering-amplitude-positron}
\overline{\mathcal{M}}=\int \frac{\mathrm{d}^4p}{(2\pi)^4}\,\delta^{(4)}(k_1+p_1-p)e^2\overline{v}^{(\alpha)}(p_1)\gamma^{\mu}\frac{-\cancel{p}+\widetilde{m}_{\psi}\mathds{1}_4}{p^2-\widetilde{m}_{\psi}^2+\mathrm{i}\epsilon}\gamma^{\nu}v^{(\alpha)}(p_1)\varepsilon^{(\lambda)}_{\mu}(k_1)\overline{\varepsilon}^{(\lambda)}_{\nu}(k_1)\,,
\end{equation}
where $\overline{\mathcal{M}}\equiv \mathcal{M}(\widetilde{\mathrm{e}}^+\upgamma\rightarrow \widetilde{\mathrm{e}}^+\upgamma)$.
Then the roles of the poles in \eqref{eq:propagator-poles} are interchanged where again the positive-energy pole is taken into account.
A similar computation to what was done before together with the sum over the spinor matrices, \eqref{eq:spinor-completeness-negative}, for the
positron spinors leads to:
\begin{align}
2\mathrm{Im}(\overline{\mathcal{M}})&=\int \frac{\mathrm{d}^3p}{(2\pi)^32\widetilde{E}_{\psi}^{(>)}}\,\delta^{(4)}(k_1+p_1-p) \notag \displaybreak[0]\\
&\phantom{{}={}\int \frac{\mathrm{d}^3p}{(2\pi)^32\widetilde{E}_{\psi}^{(>)}}}\times e^2\overline{v}^{(\alpha)}(p_1)\gamma^{\mu}\Big[\sum_{\beta=1,2} v^{(\beta)}(p)\overline{v}^{(\beta)}(p)\Big]\gamma^{\nu} v^{(\alpha)}(p_1)\varepsilon_{\mu}^{(\lambda)}(k_1)\overline{\varepsilon}_{\nu}^{(\lambda)}(k_1) \notag \displaybreak[0]\\
&=\int \frac{\mathrm{d}^3p}{(2\pi)^32\widetilde{E}_{\psi}^{(>)}}\,\delta^{(4)}(k_1+p_1-p) \notag \displaybreak[0]\\
&\phantom{{}={}\int \frac{\mathrm{d}^3p}{(2\pi)^32\widetilde{E}_{\psi}^{(>)}}}\times \sum_{\beta=1,2} \mathrm{i}e\overline{v}^{(\alpha)}(p_1)\gamma^{\mu}v^{(\beta)}(p)\varepsilon^{(\lambda)}_{\mu}(k_1)\left(\mathrm{i}e\overline{v}^{(\alpha)}(p_1)\gamma^{\nu}v^{(\beta)}(p)\varepsilon_{\nu}^{(\lambda)}(k_1)\right)^{\dagger} \notag\displaybreak[0] \\
&=\int \frac{\mathrm{d}^3p}{(2\pi)^32\widetilde{E}_{\psi}^{(>)}}\,\delta^{(4)}(k_1+p_1-p) \sum_{\beta=1,2} |\widetilde{\mathcal{M}}|^2\,,
\end{align}
with $\widetilde{\mathcal{M}}\equiv \mathcal{M}(\mathrm{\widetilde{e}^+}\upgamma \rightarrow \mathrm{\widetilde{e}^+})$.
Hence the validity of the optical theorem at tree-level can also be demonstrated for spin-1/2 antifermions. This is a
good independent crosscheck for the modified spinors, the sums over the spinor matrices, and the propagator. Since no relations were
used that only hold for the particular process considered, this proof rather general and valid for any tree-level process with
an internal electron or positron line.

A last caveat is formed by the temporal sector of \secref{sec:fermion-dispersion-laws}. The latter is characterized by the two distinct
positive-energy dispersion laws of \eqref{eq:fermion-energy-temporal} where the first of them (and its negative-energy counterpart)
is a perturbation of the standard one and the other is spurious. However the spurious solutions cannot simply be discarded when considering
the optical theorem. In this case the propagator denominator of \eqref{eq:propagator-poles} has four distinct poles and the above proof
has to be modified. Note that this issue also occurs in the context of the nonminimal {\em CPT}-even and isotropic modified Maxwell
theory \cite{Schreck:2013kja}. The problem may\footnote{For the nonminimal photon sector considered in \cite{Schreck:2013kja} the issue
appeared, if there was at least one additional time derivative. However in the context of the nonminimal fermion sector considered here
there are no spurious dispersion relations for the mixed case of $m^{(5)\alpha_1\alpha_2}$ with only one additional derivative, for example.}
occur if there are additional time
derivatives in the Dirac operator leading to an unconventional time evolution for the Dirac field (see \cite{Colladay:2001wk} for a
related problem in the minimal fermion sector). In the minimal sector it is resolved by a field redefinition at first order Lorentz
violation. This might be possible for the nonminimal case as well, but the approach introduced in \cite{Schreck:2013kja} will be employed
instead. By doing so, all additional $p^0$ components in the off-shell dispersion relation of \eqref{eq:off-shell-dispersion-law} are
replaced by the standard fermion dispersion law $p^0=\sqrt{\mathbf{p}^2+m_{\psi}^2}$: $m^{(5)00}p_0^2=m^{(5)00}(\mathbf{p}^2+m_{\psi}^2)$.
The resulting expression is then still valid at first order Lorentz violation. Computing the modified positive-energy dispersion relation after
the replacement has been performed, results in the only solution
\begin{equation}
\widetilde{E}^{(\mathrm{temp})}_{\psi'}=\sqrt{\mathbf{p}^2+\left[m_{\psi}+(\mathbf{p}^2+m_{\psi}^2)m^{(5)00}\right]^2}\,,
\end{equation}
which coincides with \eqref{eq:fermion-energy-temporal} at first order Lorentz violation. A spurious solution does not appear any more.
Then all the steps of the above proof can be redone analogously and the optical theorem at tree-level is demonstrated to be
valid at first order Lorentz violation for the temporal sector as well.

\subsection{Analysis of unitarity}
\label{sec:analysis-unitarity-mhat}

The Lorentz-violating operators involving additional time derivatives may be expected to have problems with unitarity, which will be
investigated in the current section. A useful method to study unitarity is the investigation of a property of the Euclidean propagator,
which is called reflection positivity \cite{Montvay:1994,oai:arXiv.org:hep-ph/0101087}. It states that a scalar quantum field theory obeys
a positive semi-definite self-adjoint Hamiltonian $H$ in Minkowski spacetime with a unitary time evolution if for the Euclidean two-point function
(propagator) $S_E(p^4,\mathbf{p})$ the following property is valid \cite{oai:arXiv.org:hep-ph/0101087}:
\begin{equation}
\label{eq:reflection-positivity}
\int \mathrm{d}^3p \int_{-\infty}^{\infty} \mathrm{d}p^4\,\exp(-\mathrm{i}p^4x^4)S_E(p^4,\mathbf{p})=\int \mathrm{d}^3p\,S_E(x^4,\mathbf{p})\geq 0\,.
\end{equation}
Here $p^4=-\mathrm{i}p^0$ is the Wick-rotated zeroth component of the momentum four-vector and $x^4=-\mathrm{i}x^0$ is the Wick-rotated time.
The Euclidean propagator follows from the propagator $S(p^0,\mathbf{p})$ in Minkowski spacetime via $S_E(p^4,\mathbf{p})\equiv-S(\mathrm{i}p^4,\mathbf{p})$.
The Wick-rotated propagator comes with a global minus sign, which is compensated in the latter definition (see also \cite{Klinkhamer:2010zs}).
First of all let us look at the standard quantum field theory of a scalar with mass $m_{\phi}$:
\begin{subequations}
\begin{align}
S(p^0,\mathbf{p})&=\frac{1}{p^2-m_{\phi}^2}\,, \\[2ex]
\label{eq:euclidean-propagator-standard}
S_E(p^4,\mathbf{p})&=-S(\mathrm{i}p^4,\mathbf{p})=\frac{1}{(p^4)^2+\mathbf{p}^2+m_{\phi}^2}\,.
\end{align}
\end{subequations}
To check reflection positivity the integration over $p^4$ in \eqref{eq:reflection-positivity} will be done first. This is possible by using
Eq. (3.723.2) of \cite{Gradshteyn:2007}:
\begin{align}
\label{eq:euclidean-propagator-x4}
S_E(x^4,\mathbf{p})&=\int_{-\infty}^{\infty} \mathrm{d}p^4\,\frac{\exp(-\mathrm{i}p^4x^4)}{(p^4)^2+\mathbf{p}^2+m_{\phi}^2}=2\int_0^{\infty}\mathrm{d}p^4\, \frac{\cos(p^4x^4)}{(p^4)^2+\mathbf{p}^2+m_{\phi}^2} \notag \\
&=\frac{\pi}{\sqrt{\mathbf{p}^2+m_{\phi}^2}}\exp\left(-|x^4|\sqrt{\mathbf{p}^2+m_{\phi}^2}\right)\,.
\end{align}
The latter result is manifestly positive. Hence the additional integral over the spatial momentum will be positive as well, granting reflection
positivity. The integration can also be performed explicitly by introducing spherical coordinates with $|\mathbf{p}|\equiv p$ and using Eq. (3.462.21)
of \cite{Gradshteyn:2007}:
\begin{equation}
\label{eq:reflection-positivity-standard-result}
\int \mathrm{d}^3p\,S_E(x^4,\mathbf{p})=4\pi^2 \int_0^{\infty} \mathrm{d}p\,\frac{p^2\exp\left(-|x^4|\sqrt{p^2+m_{\phi}^2}\right)}{\sqrt{p^2+m_{\phi}^2}}=4\pi^2\frac{m_{\phi}}{|x^4|}K_1(|x^4|m_{\phi})\,,
\end{equation}
where $K_1(x)$ is a particular hyperbolic Bessel function of the second kind (see Eq. (8.432) in \cite{Gradshteyn:2007}) where $K_1(x)>0$
for $x>0$.

The considerations have been performed for a scalar quantum field theory. This procedure is still justified for a Dirac theory
of spin-1/2 fermions, when omitting the matrix structure in spinor space. The reason is that the crucial information on reflection positivity
of a quantum field theory is encoded in the pole structure of the propagator. Therefore it should be sufficient to concentrate on its scalar part
and this will be done for the Lorentz-violating cases as well.

Now let us investigate reflection positivity for the scalar propagator part of the temporal case of $\widehat{m}$ with the single Lorentz-violating
coefficient $m^{(5)00}$. The scalar part of the Wick-rotated propagator reads as follows:
\begin{subequations}
\begin{align}
\label{eq:wick-rotated-propagator-temporal}
S_E(p^4,\mathbf{p})&=\frac{1}{(p^4)^2+\mathbf{p}^2+[m_{\psi}-m^{(5)00}(p^4)^2]^2}=\frac{1}{(m^{(5)00})^2}\frac{1}{[(p^4)^2+\beta_1^2][(p^4)^2+\beta_2^2]}\,, \\[2ex]
\beta_{1,2}&=\frac{\sqrt{1-2m^{(5)00}m_{\psi}\mp\sqrt{1-4(m^{(5)00})^2\mathbf{p}^2-4m^{(5)00}m_{\psi}}}}{\sqrt{2}|m^{(5)00}|}\,.
\end{align}
\end{subequations}
Now to investigate reflection positivity we have to evaluate the following integral:
\begin{equation}
S_E(x^4,\mathbf{p})=\frac{2}{(m^{(5)00})^2}\int_0^{\infty} \mathrm{d}p^4\,\frac{\cos(p^4x^4)}{[(p^4)^2+\beta_1^2][(p^4)^2+\beta_2^2]}\,,
\end{equation}
with $\beta_2$ and $\beta_1$ from above. Using Eq. (3.728.1) in \cite{Gradshteyn:2007} the analytical result reads as follows:
\begin{equation}
\label{eq:weak-reflection-positivity-case-1}
S_E(x^4,\mathbf{p})=\frac{1}{(m^{(5)00})^2}\frac{\pi}{\beta_1\beta_2}\frac{\beta_2\exp(-|x^4|\beta_1)-\beta_1\exp(-|x^4|\beta_2)}{\beta_2^2-\beta_1^2}\,.
\end{equation}
In the limit $m^{(5)00}\mapsto 0$ the standard result of \eqref{eq:euclidean-propagator-x4} is recovered. Since $\beta_{1,2}$ are positive,
the question of reflexion positivity reduces to positivity of the two-dimensional function
\begin{equation}
f(x,y)=\frac{y\exp(-ax)-x\exp(-ay)}{y-x}\,,\quad a\geq 0\,,
\end{equation}
where $(x,y)\in\mathbb{R}^+\times \mathbb{R}^+$.
For $y>x$ the denominator is larger than zero as well as the numerator since $y/x>\exp[-a(y-x)]$, which renders $f(x,y)$ positive in this case.
An analogous argument holds for $y<x$.
Hence $S_E(x^4,\mathbf{p})$ is positive for all $m^{(5)00}$. Then it can be concluded that $S_E(x^4,\mathbf{x})$ is positive as
well and reflection positivity plus unitarity is granted.

Now let us consider the mixed case with the three coefficients $m^{(5)0i}$ ($i=1\dots 3$) and the remaining ones set to zero.
Here the Wick-rotated scalar part of the propagator is
\begin{equation}
S_E(p^4,\mathbf{p})=\frac{1}{1-\widehat{m}_1^2}\frac{1}{(p^4)^2-\mathrm{i}ap^4+b^2}\,,\quad a=\frac{2\widehat{m}_1m_{\psi}}{1-\widehat{m}_1^2}\,,\quad b^2=\frac{\mathbf{p}^2+m_{\psi}^2}{1-\widehat{m}_1^2}\,,
\end{equation}
with $\widehat{m}_1$ given in \eqref{eq:fermion-energy-mixed}. Note that this result is even complex in contrast to Eqs. (\ref{eq:euclidean-propagator-standard})
and (\ref{eq:wick-rotated-propagator-temporal}). However the integral over $p^4$ is manifestly real, which is shown as follows:
\begin{align}
S_E(x^4,\mathbf{p})&=\frac{1}{1-\widehat{m}_1^2}\int_{-\infty}^{\infty} \mathrm{d}p^4\,\frac{\exp(-\mathrm{i}p^4x^4)}{(p^4)^2-\mathrm{i}ap^4+b^2} \notag \\
&=\frac{1}{1-\widehat{m}_1^2}\left[\int_0^{\infty} \mathrm{d}p^4\,\frac{\exp(-\mathrm{i}p^4x^4)}{(p^4)^2-\mathrm{i}ap^4+b^2}+\int_0^{\infty} \mathrm{d}p^4\,\frac{\exp(\mathrm{i}p^4x^4)}{(p^4)^2+\mathrm{i}ap^4+b^2}\right] \notag \\
&=\frac{2}{1-\widehat{m}_1^2}\int_0^{\infty} \mathrm{d}p^4\,\frac{[(p^4)^2+b^2]\cos(p^4x^4)+ap^4\sin(p^4x^4)}{[(p^4)^2+b^2]^2+a^2(p^4)^2}\,.
\end{align}
This integral can be computed using Eqs.~(3.728.1) -- (3.728.3) in \cite{Gradshteyn:2007}, which gives the intermediate result
\begin{subequations}
\begin{align}
S_E(x^4,\mathbf{p})&=\frac{1}{1-\widehat{m}_1^2}\frac{\pi}{\delta\varepsilon(\delta^2-\varepsilon^2)}\Big\{\delta[b^2+(\mathrm{sgn}(x^4)a-\varepsilon)\varepsilon]\exp(-|x_4|\varepsilon)\Big. \notag \\
&\phantom{{}={}\frac{1}{1-\widehat{m}_1^2}\frac{\pi}{\delta\varepsilon(\delta^2-\varepsilon^2)}\Big\{}\Big.-\varepsilon[b^2+(\mathrm{sgn}(x^4)a-\delta)\delta]\exp(-|x_4|\delta)\Big\}\,, \\[2ex]
\delta&=\frac{1}{2}\left(\sqrt{a^2+4b^2}+\mathrm{sgn}(x^4)a\right)\,,\quad \varepsilon=\frac{1}{2}\left(\sqrt{a^2+4b^2}-\mathrm{sgn}(x^4)a\right)\,,
\end{align}
\end{subequations}
with the sign function defined by \eqref{eq:sign-function}. The latter $S_E(x^4,\mathbf{p})$ can be further simplified using
\begin{subequations}
\begin{align}
&b^2+(\mathrm{sgn}(x^4)a-\varepsilon)\varepsilon=2\,\mathrm{sgn}(x^4)a\varepsilon\,,\quad b^2+(\mathrm{sgn}(x^4)a-\delta)\delta=0\,, \\[2ex]
&\delta^2-\epsilon^2=\mathrm{sgn}(x^4)a\sqrt{a^2+4b^2}\,,
\end{align}
\end{subequations}
to obtain the final amazingly short result
\begin{equation}
\label{eq:reflection-positivity-result-mixed}
S_E(x^4,\mathbf{p})=\frac{2\pi}{1-\widehat{m}_1^2}\frac{\exp(-|x_4|\varepsilon)}{\sqrt{a^2+4b^2}}\,.
\end{equation}
As long as $1-\widehat{m}_1^2>0$, which is the case for sufficiently small Lorentz-violating coefficients $m^{(5)0i}$, \eqref{eq:reflection-positivity-result-mixed}
is manifestly larger than zero. As a result, the three-dimensional integral over $S_E(x^4,\mathbf{p})$, which produces $S_E(x^4,\mathbf{x})$,
will also be larger than zero establishing reflection positivity. Note that for $m^{(5)0i}\mapsto 0$ one obtains the standard result given by
\eqref{eq:euclidean-propagator-x4}. To summarize, unitarity for the temporal and mixed sectors of $\widehat{m}$ is granted for sufficiently
small Lorentz-violating controlling coefficients.


Another quantum field theoretic property that could be studied for the nonminimal fermion sector is stability. In general this refers to the absence
of energies lying below a particular choice of ground state energy. It is known that for certain frameworks in the minimal fermion sector there exist spacelike
four-momenta with a positive energy in one particular frame \cite{Kostelecky:2000mm}. However negative energies may be generated in a sufficiently
boosted observer frame. This is why any analysis is usually restricted to a ``concordant frame,'' whose boosts are not too large. Considering the issue
of stability in the nonminimal fermion sector seems to be even more complicated than in the minimal sector. To obtain the modified fermion dispersion
laws in the current paper we have restricted ourselves to particular subsets of coefficients due to heavy computational difficulties. Now any observer
transformation may generate
additional coefficients, which leads us back to these complications. For example, applying an observer transformation in the temporal sector of
$\widehat{m}$ with the only nonzero coefficient $m^{(5)00}$ may introduce some of the mixed or spatial coefficients $m^{(5)0i}$, $m^{(5)ij}$ (for $i$, $j=1\dots 3$),
which renders the exact dispersion relations very complicated. For this reason studying the issue of stability will be postponed to future work.

\section{Application to other Lorentz-violating operators}
\label{sec:application-other-coefficients}
\setcounter{equation}{0}

In the previous sections certain properties of the quantum field theory based on the nonminimal Lorentz-violating composite operator
$\widehat{m}$ were investigated and discussed. This particular operator was chosen in the first place because it is
{\em CPT}-even and it forms a scalar under observer Lorentz transformations. Hence the corresponding parameters were supposed to be treatable
in the simplest manner. In the current section we intend to apply the considerations above to other sets of Lorentz-violating coefficients in the
nonminimal fermion sector, which have richer characteristics and may, therefore, lead to additional complications. All operators
plus their most important properties are listed and summarized in Tab.~I in \cite{Kostelecky:2013rta}.

\subsection{\textit{CPT}-even vector operator}

The first choice is the {\em CPT}-even vector operator $\widehat{c}^{\,\mu}$, which can be decomposed into a sum of
operators of even operator dimension:
\begin{equation}
\label{eq:dimensional-expansion-cmu}
\widehat{c}^{\,\mu}\equiv \widehat{c}^{\,\mu\alpha_1}p_{\alpha_1}=\sum_{\substack{d \text{ even} \\ d\geq 4}}^{\infty} \widehat{c}^{\,(d)\mu}\,,\quad \widehat{c}^{\,(d)\mu}\equiv c^{(d)\mu\alpha_1\dots \alpha_{(d-3)}}p_{\alpha_1}\dots p_{\alpha_{(d-3)}}\,,\quad \widehat{\mathcal{Q}}=\widehat{c}^{\,\mu}\gamma_{\mu}\,.
\end{equation}
The operator $\widehat{c}^{\,\mu}$ has one free Lorentz index, which makes it transform as an observer
vector by construction. Therefore it is referred to by the term ``vector operator.''
The component coefficients of the minimal dimension-4 field operator have two indices where the second is contracted with the four-momentum:
$\widehat{c}^{\,(4)\mu}=c^{(4)\mu\alpha_1}p_{\alpha_1}$. Restricting the dimensional expansion of
\eqref{eq:dimensional-expansion-cmu} to the coefficients associated with the dimension-6 field operator $\overline{\psi}\partial_{\alpha_1}\partial_{\alpha_2}\partial_{\alpha_3}\psi$
in configuration space, we deal with $\widehat{c}^{\,(6)\mu}=c^{(6)\mu\alpha_1\alpha_2\alpha_3}p_{\alpha_1}p_{\alpha_2}p_{\alpha_3}$. The
latter is made up of the 64 component coefficients $c^{(6)\mu\alpha_1\alpha_2\alpha_3}$
with mass dimension~$-2$.

\subsubsection{Modified fermion dispersion relations}
\label{sec:dispersion-relations-cmu}

In the current section the modified dispersion relations will be determined.
Equation~(35) in~\cite{Kostelecky:2013rta} gives the quantities $\widehat{\mathcal{S}}_{\pm}=-m_{\psi}$, $\widehat{\mathcal{V}}^{\mu}_{\pm}=p^{\mu}+\widehat{c}^{\,\mu}$,
and $\widehat{\mathcal{T}}^{\mu\nu}_{\pm}=0$ that are used in Eq.~(39) in the latter reference to obtain
\begin{equation}
\label{eq:off-shell-dispersion-law-chat}
(p+\widehat{c}\,)^2-m_{\psi}^2=0\,.
\end{equation}
From this polynomial the positive- and negative-energy eigenvalues can be deduced. In what follows, the positive dispersion
laws will be given. The temporal sector is characterized by the set of coefficients $c^{(6)\mu\nu00}$. This is the most complicated sector to
handle since it involves two additional time-derivatives in configuration space. The modified dispersion relations for the whole temporal
sector are involved, which is why the dispersion laws for certain subsets are given. For a theory with only a nonvanishing $c^{(6)0000}$
the modified dispersion law reads as follows:
\begin{subequations}
\label{eq:dispersion-law-temporal-cmu-1}
\begin{align}
\widetilde{E}_{\psi}^{(\mathrm{temp,1})}&=\frac{1}{\sqrt{6}}\sqrt{\frac{2^{4/3}}{\widehat{C}}+\frac{2^{2/3}}{\widehat{c}_1^{\,2}}\widehat{C}-\frac{4}{\widehat{c}_1}}\,, \\[2ex]
\widehat{C}&=\Big\{\widehat{c}_1^{\,3}\!\left[2+27(\mathbf{p}^2+m_{\psi}^2)\widehat{c}_1\right]+3\sqrt{3}\sqrt{(\mathbf{p}^2+m_{\psi}^2)\widehat{c}_1^{\,7}\left[4+27(\mathbf{p}^2+m_{\psi}^2)\widehat{c}_1\right]}\Big\}^{1/3}\,, \\[2ex]
\widehat{c}_1&=c^{(6)0000}\,.
\end{align}
\end{subequations}
Besides this perturbed dispersion law there are two further dispersion relations, which are spurious. Therefore they will not be stated here.
The occurrence of fractional powers other than square roots in \eqref{eq:dispersion-law-temporal-cmu-1} traces back to six powers of $p_0$
in \eqref{eq:off-shell-dispersion-law-chat}.

For the set of coefficients $c^{(6)0i00}$ with $i=1\dots 3$ and the remaining ones vanishing the dispersion relations are
\begin{subequations}
\label{eq:dispersion-law-temporal-cmu-2}
\begin{align}
\widetilde{E}_{\psi;1,2}^{(\mathrm{temp,2})}&=\frac{1\mp\sqrt{1-4\sqrt{\mathbf{p}^2+m_{\psi}^2}\widehat{c}_2}}{2\widehat{c}_2}\,, \\[2ex]
\widehat{c}_2&=c^{(6)0i00}p^i\,.
\end{align}
\end{subequations}
The third set of coefficients, which shall be considered for the temporal sector, is $c^{(6)ij00}$ with the spatial indices $i$ and $j$
leading to the following dispersion relations:
\begin{subequations}
\label{eq:dispersion-law-temporal-cmu-3}
\begin{align}
\widetilde{E}_{\psi}^{(\mathrm{temp,3})}&=\sqrt{\frac{1+\widehat{c}_3\mp\sqrt{(1+\widehat{c}_3)^{\,2}-4(\mathbf{p}^2+m_{\psi}^2)\widehat{c}_4}}{2\widehat{c}_4}}\,, \\[2ex]
\widehat{c}_3&=2c^{(6)ij00}p^ip^j\,,\quad \widehat{c}_4=c^{(6)ij00}c^{(6)ik00}p^jp^k\,.
\end{align}
\end{subequations}
The first of Eqs.~(\ref{eq:dispersion-law-temporal-cmu-2}) and (\ref{eq:dispersion-law-temporal-cmu-3}) are again perturbed ones and the second are
spurious. The double square root structure is specific for the dispersion relations of the temporal sector as long as the polynomial in
\eqref{eq:off-shell-dispersion-law-chat} is of degree four. The spurious dispersion laws can again be removed at first order in Lorentz violation. For the
first of the cases considered, in \eqref{eq:off-shell-dispersion-law-chat} $c^{(6)0000}p_0^2$ has to be replaced by $c^{(6)0000}(\mathbf{p}^2+m_{\psi}^2)$,
for the second $c^{(6)0i00}p_0^2$ by $c^{(6)0i00}(\mathbf{p}^2+m_{\psi}^2)$, and for the third $c^{(6)ij00}p_0^2$ by $c^{(6)ij00}(\mathbf{p}^2+m_{\psi}^2)$.
One then obtains
\begin{subequations}
\begin{align}
\label{eq:dispersion-law-1-spurious-removed}
\widetilde{E}_{\psi'}^{(\mathrm{temp,1})}&=\frac{\sqrt{\mathbf{p}^2+m_{\psi}^2}}{\left|1+(\mathbf{p}^2+m_{\psi}^2)\widehat{c}_1\right|}\,, \\[2ex]
\label{eq:dispersion-law-2-spurious-removed}
\widetilde{E}_{\psi'}^{(\mathrm{temp,2})}&=\sqrt{\mathbf{p}^2+m_{\psi}^2}\left(1+\widehat{c}_2\sqrt{\mathbf{p}^2+m_{\psi}^2}\right)\,,
\end{align}
and
\begin{equation}
\label{eq:dispersion-law-3-spurious-removed}
\widetilde{E}_{\psi'}^{(\mathrm{temp,3})}=\sqrt{(\mathbf{p}^2+m_{\psi}^2)\left[1-\widehat{c}_3+(\mathbf{p}^2+m_{\psi}^2)\widehat{c}_4\right]}\,,
\end{equation}
\end{subequations}
respectively. These are perturbed dispersion laws that coincide with the original perturbed ones (the first of Eqs.~(\ref{eq:dispersion-law-temporal-cmu-1}),
(\ref{eq:dispersion-law-temporal-cmu-2}), and (\ref{eq:dispersion-law-temporal-cmu-3}), respectively) at first order in Lorentz violation. The spurious
versions are removed by this procedure.

The mixed sector is defined by the family of component coefficients $c^{(6)\mu\nu0i}$, $c^{(6)\mu\nu i0}$ where $\mu$, $\nu$ are Lorentz indices and $i$ a spatial index.
Hence there appears one additional time derivative in configuration space in combination with these coefficients. The modified dispersion relation
associated with the whole coefficient set is involved, which is why certain subsets are considered. For nonvanishing $c^{(6)00i0}$ and $c^{(6)000i}$,
i.e., with the first two Lorentz indices set to zero one obtains
\begin{subequations}
\label{eq:dispersion-law-cmu-mixed-1}
\begin{align}
\widetilde{E}_{\psi;1,2}^{(\mathrm{mixed,1})}&=\frac{1\mp\sqrt{1-4\sqrt{\mathbf{p}^2+m_{\psi}^2}\,\widehat{c}_5}}{2\widehat{c}_5}\,, \\[2ex]
\widehat{c}_5&=(c^{(6)00i0}+c^{(6)000i})p^i\,.
\end{align}
\end{subequations}
For the latter coefficients both a perturbed and a spurious dispersion law appear again.
For $c^{(6)ijk0}$ and $c^{(6)ij0k}$, i.e., with the first two indices restricted to spatial values the modified dispersion law is given by
\begin{subequations}
\label{eq:dispersion-law-cmu-mixed-2}
\begin{align}
\widetilde{E}_{\psi}^{(\mathrm{mixed,2})}&=\frac{\widehat{c}_6+\sqrt{\widehat{c}_6^{\,2}+(\mathbf{p}^2+m_{\psi}^2)(1-\widehat{c}_7)}}{1-\widehat{c}_7}\,, \\[2ex]
\widehat{c}_6&=(c^{(6)ijk0}+c^{(6)ij0k})p^ip^jp^k\,, \\[2ex]
\widehat{c}_7&=(c^{(6)ijk0}+c^{(6)ij0k})(c^{(6)ilm0}+c^{(6)il0m})p^jp^kp^lp^m\,.
\end{align}
\end{subequations}
For this case there is only a perturbed dispersion relation, but not a spurious one.

For one of the first two indices set to zero and the remaining ones restricted to spatial values the positive-energy dispersion laws read
\begin{subequations}
\label{eq:dispersion-law-cmu-mixed-3}
\begin{align}
\widetilde{E}_{\psi}^{(\mathrm{mixed,3})}&=\sqrt{\frac{(1+\widehat{c}_8)^{\,2}+2\widehat{c}_9\mp\sqrt{\left[(1+\widehat{c}_8)^{\,2}+2\widehat{c}_9\right]^2-4(\mathbf{p}^2+m_{\psi}^2)\widehat{c}_{10}}}{2\widehat{c}_{10}}}\,, \\[2ex]
\widehat{c}_8&=(c^{(6)0i0j}+c^{(6)0ij0})p^ip^j\,, \\[2ex]
\widehat{c}_9&=(c^{(6)i0j0}+c^{(6)i00j})p^ip^j\,, \\[2ex]
\widehat{c}_{10}&=(c^{(6)i00j}+c^{(6)i0j0})(c^{(6)i00k}+c^{(6)i0k0})p^jp^k\,.
\end{align}
\end{subequations}
The spurious solutions in Eqs.~(\ref{eq:dispersion-law-cmu-mixed-1}), (\ref{eq:dispersion-law-cmu-mixed-3}) can be removed by the
replacements
\begin{subequations}
\begin{align}
&\{c^{(6)00i0},c^{(6)000i}\}p_0\mapsto \{c^{(6)00i0},c^{(6)000i}\}\sqrt{\mathbf{p}^2+m_{\psi}^2}\,, \\[2ex]
&\{c^{(6)0i0j},c^{(6)0ij0},c^{(6)i0j0},c^{(6)i00j}\}p_0\mapsto \{c^{(6)0i0j},c^{(6)0ij0},c^{(6)i0j0},c^{(6)i00j}\}\sqrt{\mathbf{p}^2+m_{\psi}^2}\,,
\end{align}
\end{subequations}
in the off-shell dispersion relation of \eqref{eq:off-shell-dispersion-law-chat}. This leads to the following perturbed dispersion laws where
the spurious versions are removed:
\begin{subequations}
\begin{align}
\label{eq:dispersion-law-4-spurious-removed}
\widetilde{E}_{\psi'}^{(\mathrm{mixed,1})}&=\sqrt{\frac{\mathbf{p}^2+m_{\psi}^2}{1+\widehat{c}_5\left[(\mathbf{p}^2+m_{\psi}^2)\widehat{c}_5-2\sqrt{\mathbf{p}^2+m_{\psi}^2}\,\right]}} \\[2ex]
\label{eq:dispersion-law-5-spurious-removed}
\widetilde{E}_{\psi'}^{(\mathrm{mixed,3})}&=\frac{\sqrt{(\mathbf{p}^2+m_{\psi}^2)\!\left\{(\widehat{c}_8+\widehat{c}_9)^2+(1-\widehat{c}_8^{\,2})\!\left[1-(\mathbf{p}^2+m_{\psi}^2)\widehat{c}_{10}\right]\!\right\}}-(\widehat{c}_8+\widehat{c}_9)\sqrt{\mathbf{p}^2+m_{\psi}^2}}{1-(\mathbf{p}^2+m_{\psi}^2)\widehat{c}_{10}}\,.
\end{align}
\end{subequations}
Finally, the spatial sector is characterized by the coefficients $c^{(6)\mu\nu ij}$ with the Lorentz indices $\mu$, $\nu$ and the
spatial indices $i$ and $j$. Due to the complexity of the general case we restrict this sector to the set of coefficients
$c^{(6)\mu ijk}$ with only one Lorentz index $\mu$ and three spatial indices $i$, $j$, and $k$. The following dispersion relation is
then associated with these coefficients:
\begin{subequations}
\label{eq:dispersion-law-cmu-spatial}
\begin{align}
\widetilde{E}_{\psi}^{(\mathrm{spatial})}&=\sqrt{\mathbf{p}^2+m_{\psi}^2+\widehat{c}_{11}}+\widehat{c}_{12}\,, \\
\widehat{c}_{11}&=c^{(6)ijkl}c^{(6)imno}p^jp^kp^lp^mp^np^o-2c^{(6)ijkl}p^ip^jp^kp^l\,, \\[2ex]
\widehat{c}_{12}&=c^{(6)0ijk}p^ip^jp^k\,.
\end{align}
\end{subequations}
Note that at least for some coefficients of the mixed and the spatial sector there are no spurious dispersion laws but only perturbed
ones.

The dispersion laws given in the current section correspond to positive energies $\widetilde{E}_{\psi}^{(>)}=\widetilde{E}_{\psi}^{(>)}(\mathbf{p},c^{(6)\mu\alpha_1\alpha_2\alpha_3})$.
The relation between the positive-energy and the negative-energy solutions $\widetilde{E}_{\psi}^{(<)}$ is $\widetilde{E}_{\psi}^{(>)}(\mathbf{p},c^{(6)\mu\alpha_1\alpha_2\alpha_3})=-\widetilde{E}_{\psi}^{(<)}(-\mathbf{p},c^{(6)\mu\alpha_1\alpha_2\alpha_3})$.
This means that both are related by reversing the sign of the four-momentum $p^{\mu}$ where the Lorentz-violating coefficients
$c^{(6)\mu\alpha_1\alpha_2\alpha_3}$ remain untouched. Also in this case the Feynman-St\"{u}ckelberg interpretation tells us that a
negative-energy particle with four-momentum $(p^{\mu})=(-p^0,-\mathbf{p})^T$ can be considered as a positive-energy antiparticle with
$(p^{\mu})=(p^0,\mathbf{p})^T$. Hence, given the particle energies $\widetilde{E}_{\psi}^{(>)}(\mathbf{p},c^{(6)\mu\alpha_1\alpha_2\alpha_3})$,
after reinterpreting $p^0=\widetilde{E}^{(<)}_{\psi}(\mathbf{p},c^{(6)\mu\alpha_1\alpha_2\alpha_3})$ with $p^{\mu}\mapsto -p^{\mu}$
the corresponding antiparticle energies result in $p^0=\widetilde{E}_{\psi}^{(>)}(\mathbf{p},c^{(6)\mu\alpha_1\alpha_2\alpha_3})$.
Then the particle and antiparticle dispersion laws are equal.

The minimal coefficients $c^{(4)\mu\alpha_1}$, which are linked to the dimension-4 field operator, are both {\em CPT}-even and {\em C}-even
(see \cite{Kostelecky:2008ts} for the transformation properties of the various Lorentz-violating operators with respect to {\em C}, {\em P},
and {\em T}). For this reason the $c^{(4)\mu\alpha_1}$ in the positive-energy solutions do not come with a different sign in comparison to
the negative-energy solutions \cite{Kostelecky:2000mm}. The same holds for the coefficients $c^{(6)\mu\alpha_1\alpha_2\alpha_3}$,
which substantiates the computed results.

However, caution is required when talking about $\widehat{c}^{\,\mu}$, which includes additional, contracted four-derivatives in configuration
space. For example, the dimension-4 coefficients are contracted with one four-derivative $\partial_{\alpha_1}$. A four-derivative transforms odd
under {\em CPT}, whereby $c^{(4)\mu\alpha_1}\partial_{\alpha_1}$ is {\em CPT}-odd as well. Hence based on the {\em CPT}-handedness of
the minimal coefficients the transformation properties of the coefficients contracted with additional four-derivatives depends on the number of
these derivatives. Therefore the nonminimal dimension-6 coefficients contracted with three derivatives, $c^{(6)\mu\alpha_1\alpha_2\alpha_3}\partial_{\alpha_1}\partial_{\alpha_2}\partial_{\alpha_3}$,
transform as a {\em CPT}-odd object. Similar arguments are valid in momentum space. This is why for the
antiparticle energies of Eq. (65) in \cite{Kostelecky:2013rta} the sign in the second term is different from the sign
of the particle energies of Eq. (61).

The expansions of Eqs.~(\ref{eq:dispersion-law-1-spurious-removed}) -- (\ref{eq:dispersion-law-cmu-spatial}) at first
order in Lorentz violation agree with the upper $2\times 2$ block of Eq.~(59) in \cite{Kostelecky:2013rta} and the results
for antiparticles agree with the reinterpreted lower $2\times 2$ block of the latter equation.

\subsubsection{Effective operators}

Certain operators in the fermion sector are related, e.g., $\widehat{m}$ and $\widehat{c}^{\,\mu}$ \cite{Kostelecky:2013rta}.
For example, expanding the dispersion relation of \eqref{eq:dispersion-law-temporal-cmu-1} for the temporal sector of
$\widehat{c}^{\,\mu}$ the following result is obtained at first order in the single nonzero Lorentz-violating coefficient:
\begin{equation}
\label{eq:effective-coefficients-dispersion-mhat-1}
\widetilde{E}_{\psi;\widehat{c}^{\,0}}^{(\mathrm{temp})}=E_{\psi}-c^{(6)0000}E_{\psi}^3=E_{\psi}-\widehat{c}^{\,0}\,,
\end{equation}
with the standard fermion energy $E_{\psi}$. Compare this result to the first-order expansion of the dispersion relation of
\eqref{eq:fermion-energy-temporal-perturbed},
\begin{equation}
\label{eq:effective-coefficients-dispersion-cmu-1}
\widetilde{E}_{\psi;\widehat{m}}^{(\mathrm{temp})}=E_{\psi}+m_{\psi}m^{(5)00}E_{\psi}=E_{\psi}+\frac{m_{\psi}}{E_{\psi}}\widehat{m}\,,
\end{equation}
which is valid for the temporal sector of $\widehat{m}$. They have a similar structure, i.e., the respective Lorentz-violating
operators may be related to each other. For this reason an effective operator can be introduced that incorporates both
the $\widehat{m}$ and $\widehat{c}^{\,\mu}$ operator. Since $\widehat{c}^{\,\mu}$ transforms as a Lorentz vector and
$\widehat{m}$ as a Lorentz scalar, the following \textit{Ansatz} is proposed for the effective operator:
\begin{equation}
\widehat{c}^{\,\mu}_{\mathrm{eff}}=\alpha \widehat{c}^{\,\mu}+\beta p^{\mu} \widehat{m}\,,
\end{equation}
where the four-momentum $p^{\mu}$ is used to provide $\widehat{m}$ a vector structure. Now the parameters $\alpha$ and
$\beta\in \mathbb{R}$ have to be determined. By contracting the \textit{Ansatz} above with $-p_{\mu}/E_{\psi}$ and setting
$\widehat{c}^{\,i}=0$ (for $i=1\dots 3$) we try to reproduce the first-order terms in the dispersion laws:
\begin{equation}
-\frac{1}{E_{\psi}}p_{\mu}\widehat{c}^{\,\mu}_{\mathrm{eff}}=-\alpha \widehat{c}^{\,0}-\beta \frac{m_{\psi}^2}{E_{\psi}}\widehat{m}\,.
\end{equation}
Comparing this with Eqs.~(\ref{eq:effective-coefficients-dispersion-mhat-1}) and (\ref{eq:effective-coefficients-dispersion-cmu-1}),
respectively, delivers $\alpha=1$ and $\beta=-1/m_{\psi}$. Hence the effective operator would be given by
\begin{equation}
\widehat{c}^{\,\mu}_{\mathrm{eff}}=\widehat{c}^{\,\mu}-\frac{1}{m_{\psi}}p^{\mu} \widehat{m}\,.
\end{equation}
This is in accordance with the second equation of Eqs.~(26) in \cite{Kostelecky:2013rta}. Now let us look at the zeroth
component:
\begin{equation}
\widehat{c}^{\,0}_{\mathrm{eff}}\equiv \widehat{c}^{\,0}-\frac{E_{\psi}}{m_{\psi}}\widehat{m}\,,\quad c^{(6)0000}_{\mathrm{eff}}=c^{(6)0000}-\frac{1}{m_{\psi}}m^{(5)00}\,.
\end{equation}
The latter result coincides with the second equation of Eqs.~(27) in \cite{Kostelecky:2013rta} for $d=6$. The next step is
to compare the expansions of the dispersion relations for the spatial sector of Eqs. (\ref{eq:fermion-energy-spatial-expanded})
and (\ref{eq:dispersion-law-cmu-spatial}):
\begin{subequations}
\begin{align}
\label{eq:effective-coefficients-dispersion-mhat-2}
\widetilde{E}_{\psi;\widehat{m}}^{(\mathrm{spat})}&=E_{\psi}+\frac{m_{\psi}}{E_{\psi}}m^{(5)kl}p^kp^l=E_{\psi}+\frac{m_{\psi}}{E_{\psi}}\widehat{m}\,, \\[2ex]
\label{eq:effective-coefficients-dispersion-pmu-2}
\widetilde{E}_{\psi;\widehat{c}^{\,i}}^{(\mathrm{spat})}&=E_{\psi}-\frac{1}{E_{\psi}}c^{(6)ijkl}p^ip^jp^kp^l=E_{\psi}+\frac{1}{E_{\psi}}\widehat{c}^{\,i}p^i\,.
\end{align}
\end{subequations}
Repeating the procedure above, i.e., contracting the \textit{Ansatz} for the effective operator with $-p_{\mu}/E_{\psi}$
and setting $\widehat{c}^{\,0}=0$ results in
\begin{equation}
-\frac{1}{E_{\psi}}p_{\mu}\widehat{c}^{\,\mu}_{\mathrm{eff}}=\frac{1}{E_{\psi}}\alpha p^i\widehat{c}^{\,i}-\beta \frac{m_{\psi}^2}{E_{\psi}}\widehat{m}\,,
\end{equation}
from which $\alpha=1$ and $\beta=-1/m_{\psi}$ follows when it is compared to Eqs.~(\ref{eq:effective-coefficients-dispersion-mhat-2})
and (\ref{eq:effective-coefficients-dispersion-pmu-2}). Hence the \textit{Ansatz} is consistent for both sectors. Considering the
$i$-th component of $\widehat{c}^{\,\mu}_{\mathrm{eff}}$ and multiplying it with $p^i$ leads to
\begin{equation}
\widehat{c}^{\,i}_{\mathrm{eff}}p^i\equiv \widehat{c}^{\,i}p^i-\frac{1}{m_{\psi}}\delta^{ij}p^ip^j\widehat{m}\,,\quad c^{(6)ijkl}_{\mathrm{eff}}=c^{(6)ijkl}+\frac{1}{m_{\psi}}\delta^{ij}m^{(5)kl}\,.
\end{equation}
which is again in accordance with the second equation of Eqs.~(27) in \cite{Kostelecky:2013rta} for $d=6$. Similar deliberations can be done for
the other coefficients. This provides a good cross check for the results obtained.

\subsubsection{Modified spinors and sums over spinor matrices}

On the one hand, according to \cite{Kostelecky:2013rta} the positive-energy spinors can be written as follows:
\begin{equation}
u^{(\alpha)}(\widetilde{E}_{\psi}^{(>)},\mathbf{p})=\frac{1}{\sqrt{N_u^{(\alpha)}}}U^{\dagger}(\widetilde{E}_{\psi}^{(>)}+\widehat{c}^{\,0},m_{\psi},\mathbf{p}+\widehat{\mathbf{c}})u^{(\alpha)}(m_{\psi},\mathbf{0})\,,\quad \widehat{c}^{\,\mu}=\widehat{c}^{\,\mu}(\widetilde{E}_{\psi}^{(>)},\mathbf{p})\,,
\end{equation}
where the $u^{(\alpha)}(m_{\psi},\mathbf{0})$ are given in \eqref{eq:positive-energy-spinors-at-rest}. These spinors are a solution of
the modified Dirac equation $(\cancel{p}+\cancel{\widehat{c}}-m_{\psi}\mathds{1}_4)u^{(\alpha)}(p)=0$ with $(p^{\mu})=(\widetilde{E}_{\psi}^{(>)},\mathbf{p})^T$.
The modified Dirac equation and the spinor solution show that $\widehat{c}^{\,\mu}$ is tightly connected to the particle four-momentum.
On the other hand, the negative-energy spinors are given by:
\begin{subequations}
\begin{align}
v^{(\alpha)}(\widetilde{E}_{\psi}^{(>)},\mathbf{p})&=\frac{1}{\sqrt{N_v^{(\alpha)}}}U^{\dagger}(\widetilde{E}_{\psi}^{(>)}+\widehat{c}^{\,0},m_{\psi},-\mathbf{p}-\widehat{\mathbf{c}})v^{(\alpha)}(m_{\psi},\mathbf{0})\,, \\[2ex]
\label{eq:property-cmu}
\widehat{c}^{\,\mu}&=\widehat{c}^{\,\mu}(-\widetilde{E}_{\psi}^{(>)},-\mathbf{p})=-\widehat{c}^{\,\mu}(\widetilde{E}_{\psi}^{(>)},\mathbf{p})\,,
\end{align}
\end{subequations}
with the $v^{(\alpha)}(m_{\psi},\mathbf{0})$ of \eqref{eq:negative-energy-spinors-at-rest}. The property (\ref{eq:property-cmu}) of the
$\widehat{c}^{\,\mu}$ operator is valid since it contains a combination of three four-momenta. These spinors obey the modified Dirac equation
$(\cancel{p}+\cancel{\widehat{c}}-m_{\psi}\mathds{1}_4)v^{(\alpha)}(p)=0$ with $(p^{\mu})=(-\widetilde{E}_{\psi}^{(>)},-\mathbf{p})^T$.
The normalizations $N_u^{(\alpha)}$ and $N_v^{(\alpha)}$ of the positive- and negative-energy spinors are chosen such that
\begin{equation}
\label{eq:spinor-normalization-conditions-cmu}
\overline{u}^{(\alpha)}(p)u^{(\beta)}(p)=2m_{\psi}\delta^{\alpha\beta}\,,\quad \overline{v}^{(\alpha)}(p)v^{(\beta)}(p)=-2m_{\psi}\delta^{\alpha\beta}\,.
\end{equation}
The sums over the positive- and negative-energy spinor matrices are given by:
\begin{subequations}
\label{eq:spinor-completeness-cmu}
\begin{align}
\sum_{\alpha=1,2} u^{(\alpha)}(p)\overline{u}^{(\alpha)}(p)&=\cancel{p}+\cancel{\widehat{c}}+m_{\psi}\mathds{1}_4\,, \\[2ex]
\sum_{\alpha=1,2} v^{(\alpha)}(p)\overline{v}^{(\alpha)}(p)&=\cancel{p}+\cancel{\widehat{c}}-m_{\psi} \mathds{1}_4\,,
\end{align}
\end{subequations}
where $(p^{\mu})=(\widetilde{E}_{\psi}^{(>)},\mathbf{p})^T$ on the right-hand sides of the latter two equations.
All these relations can be shown analogously to the relations for the operator $\widehat{m}$, cf. Appx.~\ref{sec:spinors-completeness-explicit-mmu},
\ref{sec:spinors-completeness-explicit-cmu}, which again indicates that $\widehat{m}$ and $\widehat{c}^{\,\mu}$ are related. With the
propagator
\begin{equation}
S(p)=\frac{\mathrm{i}}{(p+\widehat{c}\,)^2-m_{\psi}^2}\left(\cancel{p}+\cancel{\widehat{c}}+m_{\psi}\mathds{1}_4\right)\,,
\end{equation}
the proof of the optical theorem for the process considered in the last chapter can be done completely analogously. Note that in
the forward scattering amplitude for the positron given by \eqref{eq:forward-scattering-amplitude-positron} the sign of the
four-momentum vector has to be reversed where, as a result of this, the sign of $\widehat{c}$ changes as well. This leads to
the second relation of \eqref{eq:spinor-completeness-cmu}.
Furthermore for the check of the optical theorem for sets of component coefficients with spurious dispersion relations their
replacements, which are valid at first order Lorentz violation, have to be used (e.g., Eqs.~(\ref{eq:dispersion-law-1-spurious-removed}) --
(\ref{eq:dispersion-law-3-spurious-removed}) for the temporal sector and Eqs.~(\ref{eq:dispersion-law-4-spurious-removed}),
(\ref{eq:dispersion-law-5-spurious-removed}) for the mixed sector of the operator $\widehat{c}^{\,\mu}$).

\subsubsection{Analysis of unitarity}
\label{sec:analysis-unitarity-chatmu}

The vector operator $\widehat{c}^{\,\mu}$ also involves sectors with additional time derivatives, which may be
expected to cause issues with unitarity. Therefore the property of reflection positivity shall be studied for a particular subset of
coefficients such as it was done for the temporal and mixed sector of $\widehat{m}$ in \secref{sec:analysis-unitarity-mhat}. We will
do this for the case of the single component coefficient $c^{(6)0000}$ and all others set to zero. The Wick-rotated scalar part of the
propagator is given by
\begin{equation}
S_E(p^4,\mathbf{p})=\frac{1}{(p^4)^2[1-(p^4)^2c^{(6)0000}]^2+\mathbf{p}^2+m_{\psi}^2}\,.
\end{equation}
Now the first step to show reflection positivity is to compute the integral
\begin{align}
S_E(x^4,\mathbf{p})&=\int_{-\infty}^{\infty} \mathrm{d}p^4 \exp(-\mathrm{i}p^4x^4)S_E(p^4,\mathbf{p}) \notag \\
&=2\int_0^{\infty} \mathrm{d}p^4\,\frac{\cos(p^4x^4)}{(p^4)^2[1-(p^4)^2c^{(6)0000}]^2+\mathbf{p}^2+m_{\psi}^2}\,.
\end{align}
The denominator of the integrand involves a polynomial of sixth degree in $p^4$, which makes the computation of the integral quite
involved. However it is possible to perform some general and also numerical analyses.

For small $p^4$, which means that $(p^4)^2c^{(6)0000}\ll 1$,
the denominator resembles the standard case $(p^4)^2+\mathbf{p}^2+m_{\psi}^2$, whereby the integrand is only slightly modified.
However there are regions of $p^4$ where the structure of the denominator is highly modified compared to the standard
case. There always exists a particular intermediate $(p^4)_{\mathrm{int}}$ for which $1-(p^4)^2c^{(6)0000}$ vanishes. In the neighborhood
of $(p^4)_{\mathrm{int}}$ the integrand strongly differs from the standard result. Last but not least, for very large $p^4$, i.e., $(p^4)^2c^{(6)0000}\gg 1$,
the quartic term in $p^4$ is dominant, which heavily suppresses the integrand in comparison to the standard one.
\begin{figure}[b!]
\subfloat[]{\label{fig:plots-integrand-reflection-positivity-1}\includegraphics[scale=0.8]{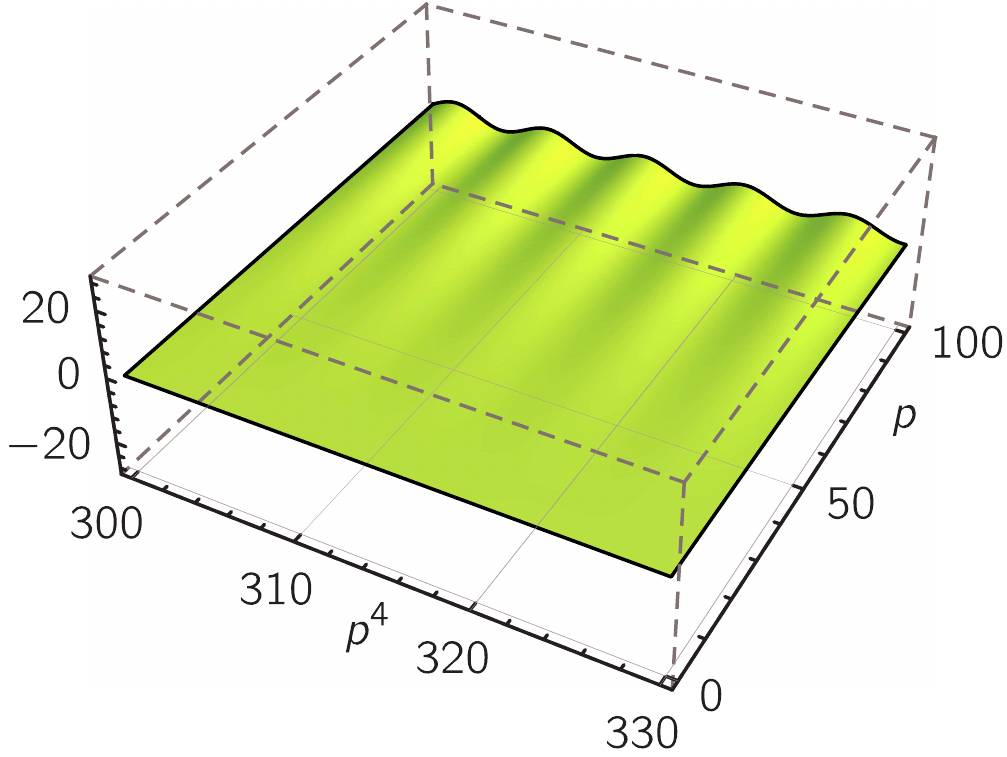}}
\subfloat[]{\label{fig:plots-integrand-reflection-positivity-2}\includegraphics[scale=0.8]{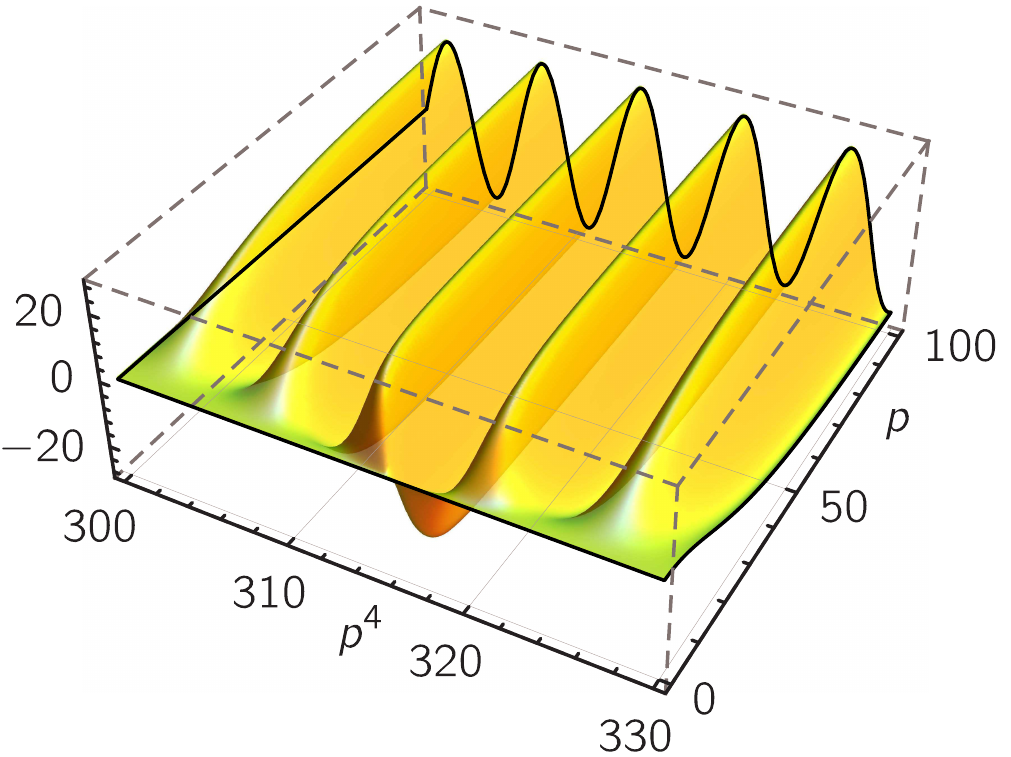}}
\caption{Plots of the integrand of \eqref{eq:reflection-positivity-temporal-cmu} as a function of $p^4$ and $p$ with the choices $m_{\psi}=1$,
$x^4=1$ for distinct values of the Lorentz-violating component coefficient; (a) for $c^{(6)0000}=0$ and (b) for $c^{(6)0000}=10^{-5}$.}
\label{fig:plots-integrand-reflection-positivity}
\end{figure}

Due to the region around $(p^4)_{\mathrm{int}}$ it is not clear whether reflection positivity can still be granted even for a small
Lorentz-violating coefficient. Since the Euclidean propagator is isotropic, spherical coordinates can be introduced with $p\equiv |\mathbf{p}|$.
This leads to the following integral:
\begin{equation}
\label{eq:reflection-positivity-temporal-cmu}
S_E(x^4,\mathbf{x})=8\pi \int_0^{\infty} \mathrm{d}p\,p^2 \int_0^{\infty} \mathrm{d}p^4\,\frac{\cos(p^4x^4)}{(p^4)^2[1-(p^4)^2c^{(6)0000}]^2+p^2+m_{\psi}^2}\,.
\end{equation}
In \figref{fig:plots-integrand-reflection-positivity-1} the integrand of $S_E(x^4,\mathbf{x})$ is plotted for a vanishing Lorentz-violating component
coefficient and in \figref{fig:plots-integrand-reflection-positivity-2} for a nonzero value much smaller than 1. This was done to illustrate that there exist
regions for which the integrand is strongly modified compared to the standard case even for a small Lorentz-violating coefficient. The modification
in \figref{fig:plots-integrand-reflection-positivity-2} is not simply proportional to $c^{(6)0000}$, but it may exceed the value of the coefficient by
several orders of magnitude.

A numerical integration of \eqref{eq:reflection-positivity-temporal-cmu} is complicated as well because of the highly oscillatory integrand
for large $p^4$. Cutting off the integration over $p$ at $\Lambda=10^2$ seems to help where it was checked that the result does not change significantly
by using a larger cut-off. For special values of $x^4$, $m_{\psi}$, and the Lorentz-violating coefficient we obtain the numerical results shown in
\tabref{tab:reflection-positivity-temporal-cmu-numerical}.
\begin{table}[t!]
\begin{tabular}{cr}
\toprule
$c^{(6)0000}$ & $S_E(x^4,\mathbf{x}$) \\
\colrule
$10^{-6}$ & $96.9042$ \\
$10^{-5}$ & $-40.0508$ \\
$10^{-4}$ & $139.043$ \\
$10^{-3}$ & $148.369$ \\
\botrule
\end{tabular}
\caption{The current table presents numerical results for \eqref{eq:reflection-positivity-temporal-cmu} for the special values
$x^4=1$ and $m_{\psi}=1$. The particular Lorentz-violating coefficient chosen is shown in the first column where the second column
gives the integration result.}
\label{tab:reflection-positivity-temporal-cmu-numerical}
\end{table}

Hence there exist particular choices of the Lorentz-violating coefficient, for which reflection positivity is violated. This also
indicates a possible violation of unitarity and it is an interesting result, which must be substantiated in future research. Note that such a behavior
does not occur for the scalar operator $\widehat{m}$. In fact, for the latter sector $p^4$ can be chosen such that the mass term in the
denominator of the propagator vanishes, cf. \eqref{eq:wick-rotated-propagator-temporal}. However this does not drastically modify the $p^4$-dependence.
Only for very large $p^4$, i.e., $(p^4)^2m^{(5)00}\gg 1$ the integrand is also suppressed compared to the standard case. But this does not
seem to violate reflection positivity as was shown in \secref{sec:analysis-unitarity-mhat}.

\subsection{\textit{CPT}-odd pseudoscalar operator}

Only for the Lorentz-violating operators previously considered, the diagonalization of the Dirac operator $\cancel{p}-m_{\psi}\mathds{1}_4+\widehat{\mathcal{Q}}$
can be performed with the matrix $U$ given by \eqref{eq:diagonalization-matrix}. For all remaining cases diagonalization is more involved
and an analogue of $U$ valid at all orders in Lorentz violation is not at hand so far. However it was shown that in these cases the Dirac operator
can be block-diagonalized at least at first order Lorentz violation by the following matrix \cite{Kostelecky:2013rta}:
\begin{equation}
\label{eq:diagonalization-matrix-first-order}
U^{(1)}=\left(\mathds{1}_4+\frac{1}{4E_{\psi}}[\gamma^5,R]\right)VW\,,\quad R=VW\gamma^0\widehat{\mathcal{Q}}W^{\dagger}V^{\dagger}\,,\quad W=W(E_{\psi},m_{\psi},\mathbf{p})\,,
\end{equation}
with $V$ and $W$ of \eqref{eq:diagonalization-matrix}. Here the index of $U$ indicates that this result is valid at first order in Lorentz
violation. Note that the expression involves the standard fermion dispersion relation $E_{\psi}$ (except of in $\widehat{\mathcal{Q}}$).

As a next example the {\em CPT}-odd observer pseudoscalar operator $\widehat{f}\equiv \widehat{f}^{\alpha_1}p_{\alpha_1}$ will be
considered. It can be written as a series of composite operators with even operator dimension:
\begin{equation}
\widehat{f}\equiv \widehat{f}^{\alpha_1}p_{\alpha_1}=\sum_{\substack{d \text{ even} \\ d\geq 4}}^{\infty} \widehat{f}^{(d)}\,,\quad \widehat{f}^{(d)}\equiv f^{(d)\alpha_1\dots \alpha_{(d-3)}}p_{\alpha_1}\dots p_{\alpha_{(d-3)}}\,,\quad \widehat{\mathcal{Q}}=\mathrm{i}\widehat{f}\gamma^5\,.
\end{equation}
First of all, $\widehat{f}$ does not have a free Lorentz index, which makes it transform as a scalar under proper observer Lorentz
transformations. Due to the $\gamma^5$-matrix a parity-transformation gives an additional sign, which is why this operator is denoted as an observer
``pseudoscalar.''
The minimal extension comprises the dimension-4 operator with $\widehat{f}^{(4)}=f^{(4)\alpha_1}p_{\alpha_1}$ where the corresponding
coefficients $f^{(4)\alpha_1}$ are contracted with the four-momentum. In this particular section the composite operator
$\widehat{f}^{(6)}=f^{(6)\alpha_1\alpha_2\alpha_3}p_{\alpha_1}p_{\alpha_2}p_{\alpha_3}$ will be considered, which is associated to the
dimension-6 field operator $\overline{\psi}\partial_{\alpha_1}\partial_{\alpha_2}\partial_{\alpha_3}\psi$.
The 20 component coefficients $f^{(6)\alpha_1\alpha_2\alpha_3}$ have mass dimension $-2$. We again split this
set of component coefficients into a temporal, mixed, and a spatial sector. The temporal sector consists of the single coefficient $f^{(6)\alpha_100}$, the mixed
sector is made up of $f^{(6)\alpha_10i}$, $f^{(6)\alpha_1i0}$ with the spatial index $i$ and the spatial sector comprises $f^{(6)\alpha_1ij}$ with
$i$, $j=1\dots 3$.

Note that care has to be taken when talking about the {\em CPT}-handedness of the operator $\widehat{f}$. The
corresponding operator in the Lagrange density is {\em CPT}-odd, but this property may be fictitious. That will be elaborated
on in \secref{sec:connection-cmu-f}. Nevertheless referring to Tab.~I of \cite{Kostelecky:2013rta}, the operator $\widehat{f}$
will be called {\em CPT}-odd within the current article.

\subsubsection{Modified dispersion relations}
\label{sec:dispersion-laws-fmu}

The modified positive-energy dispersion laws $\widetilde{E}_{\psi}^{(>)}$ follow from the condition $\det(\cancel{p}-m_{\psi}\mathds{1}_4+\widehat{\mathcal{Q}})=0$
with $\widehat{\mathcal{Q}}=\mathrm{i}\widehat{f}\gamma^5$. An alternative is to use Eq.~(39) of \cite{Kostelecky:2013rta} with
$\widehat{\mathcal{S}}_{\pm}=-m_{\psi}\pm\mathrm{i}\widehat{f}$, $\widehat{\mathcal{V}}^{\mu}_{\pm}=p^{\mu}$, and $\widehat{\mathcal{T}}^{\mu\nu}_{\pm}=0$.
For the temporal sector with the single nonvanishing coefficient $f^{(6)000}$ there are two distinct positive energies. The first reads
\begin{subequations}
\label{eq:dispersion-temporal-fmu-1}
\begin{align}
\widetilde{E}_{\psi;1}^{(\mathrm{temp,1})}&=\frac{1}{3^{1/4}}\sqrt{\frac{2}{|\widehat{f}_1|}\sin\left(\frac{u}{3}+\frac{\pi}{6}\right)}\,, \\[2ex]
u&=-\arctan\left(\frac{\sqrt{12-81(\mathbf{p}^2+m_{\psi}^2)^2(\widehat{f}_1)^2}}{9(\mathbf{p}^2+m_{\psi}^2)|\widehat{f}_1|}\right)\,, \\[2ex]
\widehat{f}_1&=f^{(6)000}\,,
\end{align}
\end{subequations}
and the second is given by
\begin{subequations}
\label{eq:dispersion-temporal-fmu-1-spurious}
\begin{align}
\widetilde{E}_{\psi;2}^{(\mathrm{temp,1})}&=\frac{1}{6^{1/3}}\sqrt{\frac{2\cdot 3^{1/3}}{v^{1/3}}+\frac{(2v)^{1/3}}{(\widehat{f}_1)^2}}\,, \\[2ex]
v&=\sqrt{81(\mathbf{p}^2+m_{\psi}^2)^2(\widehat{f}_1)^8-12(\widehat{f}_1)^6}-9(\mathbf{p}^2+m_{\psi}^2)(\widehat{f}_1)^4\,.
\end{align}
\end{subequations}
These involve trigonometric functions and fractional powers other than square roots, which are rather unusual functions to appear in
the context of modified dispersion laws. The reason for their occurrence is that the modified determinant condition is a polynomial
of sixth degree in $p^0$.\footnote{Dispersion relations involving trigonometric functions and fractional powers different from mere
square roots also appear in the mixed sector of the dimension-6 coefficients $\kappa_{\mathrm{tr}-}^{\mu\nu}$ in the nonminimal SME
photon sector \cite{Schreck:2013kja}.}
The first dispersion law is perturbed and the second is spurious, which becomes evident from the following expansions in the Lorentz-violating
coefficient:
\begin{subequations}
\begin{align}
\widetilde{E}_{\psi;1}^{(\mathrm{temp,1})}&=\sqrt{\mathbf{p}^2+m_{\psi}^2}\left[1+\frac{1}{2}(\widehat{f}_1)^2(\mathbf{p}^2+m_{\psi}^2)^2\right]+\mathcal{O}[(\widehat{f}_1)^4]\,, \\[2ex]
\widetilde{E}_{\psi;2}^{(\mathrm{temp,1})}&=\frac{1}{\sqrt{\widehat{f}_1}}-\frac{1}{4}\sqrt{\widehat{f}_1}(\mathbf{p}^2+m_{\psi}^2)+\mathcal{O}[(\widehat{f}_1)^{3/2}]\,.
\end{align}
\end{subequations}
Considering the temporal sector with the three coefficients $f^{(6)i00}$ for $i=1\dots 3$ and all remaining ones set to zero the positive-energy
dispersion laws are
\begin{subequations}
\label{eq:dispersion-temporal-fmu-2}
\begin{align}
\widetilde{E}_{\psi;1,2}^{(\mathrm{temp,2})}&=\frac{\sqrt{1\mp\sqrt{1-4(\widehat{f}_2)^2(\mathbf{p}^2+m_{\psi}^2)}}}{\sqrt{2}|\widehat{f}_2|}\,, \\[2ex]
\widehat{f}_2&=f^{(6)i00}p^i\,.
\end{align}
\end{subequations}
Here the first is perturbed and the second spurious. The double square root structure appears again since the determinant condition
is a polynomial of fourth degree in $p^0$. The spurious dispersion laws of Eqs.~(\ref{eq:dispersion-temporal-fmu-1-spurious}),
(\ref{eq:dispersion-temporal-fmu-2}) can be removed at first order Lorentz violation by the replacement
$f^{(6)\mu00}p_0^2\mapsto f^{(6)\mu00}(\mathbf{p}^2+m_{\psi}^2)$ in the determinant condition. This leads to
\begin{subequations}
\begin{align}
\label{eq:dispersion-law-fmu-1-spurious-removed-1}
\widetilde{E}_{\psi'}^{(\mathrm{temp,1})}&=\sqrt{\frac{\mathbf{p}^2+m_{\psi}^2}{1-(\mathbf{p}^2+m_{\psi}^2)^2(\widehat{f}_1)^2}} \\[2ex]
\label{eq:dispersion-law-fmu-1-spurious-removed-2}
\widetilde{E}_{\psi'}^{(\mathrm{temp,2})}&=\sqrt{(\mathbf{p}^2+m_{\psi}^2)\left[1+(\mathbf{p}^2+m_{\psi}^2)(\widehat{f}_2)^2\right]}\,.
\end{align}
\end{subequations}
For the mixed sector first of all, the coefficients are considered with the first Lorentz index equal to zero. The positive-energy
dispersion laws read
\begin{subequations}
\label{eq:dispersion-mixed-fmu-1}
\begin{align}
\widetilde{E}_{\psi;1,2}^{(\mathrm{mixed,1})}&=\frac{\sqrt{1\mp\sqrt{1-4(\widehat{f}_3)^2(\mathbf{p}^2+m_{\psi}^2)}}}{\sqrt{2}|\widehat{f}_3|}\,, \\[2ex]
\widehat{f}_3&=(f^{(6)0i0}+f^{(6)00i})p^i\,.
\end{align}
\end{subequations}
Here the first is perturbed and the second is spurious. Equations~(\ref{eq:dispersion-temporal-fmu-2}) and (\ref{eq:dispersion-mixed-fmu-1})
can, in principle, be merged into a single dispersion relation dependent on the combination of the two sets of coefficients.
To remove the spurious dispersion law in \eqref{eq:dispersion-mixed-fmu-1} the replacements
$\{f^{(6)0i0},f^{(6)00i}\}p_0\mapsto \{f^{(6)0i0},f^{(6)00i}\}\sqrt{\mathbf{p}^2+m_{\psi}^2}$ in the determinant condition have to be performed,
which lead to
\begin{equation}
\label{eq:dispersion-law-fmu-1-spurious-removed-3}
\widetilde{E}_{\psi'}^{(\mathrm{mixed,1})}=\sqrt{\frac{\mathbf{p}^2+m_{\psi}^2}{1-(\mathbf{p}^2+m_{\psi}^2)(\widehat{f}_3)^2}}\,.
\end{equation}
Note the similarities between Eqs.~(\ref{eq:dispersion-law-fmu-1-spurious-removed-1}) and (\ref{eq:dispersion-law-fmu-1-spurious-removed-3}).

Second, only the coefficients where the first Lorentz index has a spatial value are taken into account. For this particular set one then obtains
\begin{subequations}
\label{eq:dispersion-mixed-fmu-2}
\begin{align}
\widetilde{E}_{\psi}^{(\mathrm{mixed,2})}&=\sqrt{\frac{\mathbf{p}^2+m_{\psi}^2}{1-(\widehat{f}_4)^2}}\,, \\[2ex]
\widehat{f}_4&=(f^{(6)ij0}+f^{(6)i0j})p^ip^j\,.
\end{align}
\end{subequations}
Finally, for the spatial sector it follows that
\begin{subequations}
\label{eq:dispersion-spatial-fmu}
\begin{align}
\widetilde{E}_{\psi}^{(\mathrm{spatial})}&=\frac{-\widehat{f}_5\widehat{f}_6+\sqrt{[\mathbf{p}^2+m_{\psi}^2+(\widehat{f}_6)^2][1-(\widehat{f}_5)^2]+(\widehat{f}_5\widehat{f}_6)^2}}{1-(\widehat{f}_5)^2}\,, \\[2ex]
\widehat{f}_5&=f^{(6)0ij}p^ip^j\,,\quad \widehat{f}_6=f^{(6)ijk}p^ip^jp^k\,.
\end{align}
\end{subequations}
For at least some coefficients of the spatial and the mixed sector there are only perturbed but no spurious dispersion laws. The
negative-energy solutions are related to be positive-energy solutions by
$\widetilde{E}_{\psi}^{(>)}(\mathbf{p},f^{(6)\alpha_1\alpha_2\alpha_3})=-\widetilde{E}_{\psi}^{(<)}(-\mathbf{p},-f^{(6)\alpha_1\alpha_2\alpha_3})$.
Note that contrary to the cases with the $\widehat{m}$ and the $\widehat{c}^{\,\mu}$ operators, the $f^{(6)\alpha_1\alpha_2\alpha_3}$ come
with a minus sign on the right-hand side of the latter relation. This indicates their {\em CPT}-odd nature and the same behavior is observed
for the minimal, {\em CPT}-odd SME coefficients in the fermion sector \cite{Kostelecky:2000mm}. So if $p^0=\widetilde{E}_{\psi}^{(<)}(\mathbf{p},f^{(6)\alpha_1\alpha_2\alpha_3})$
is reinterpreted with the transformation $p^{\mu} \mapsto -p^{\mu}$, the antiparticle energies will be $p^0=\widetilde{E}_{\psi}^{(>)}(\mathbf{p},-f^{(6)\alpha_1\alpha_2\alpha_3})$.
The latter differ from the corresponding particle energies $\widetilde{E}_{\psi}^{(>)}(\mathbf{p},f^{(6)\alpha_1\alpha_2\alpha_3})$ due to the minus
sign associated with the $f^{(6)\alpha_1\alpha_2\alpha_3}$. This is expected for a theory violating {\em CPT}.

Furthermore, in contrast to the cases of the operators $\widehat{m}$ and $\widehat{c}$ the Lorentz-violating operator $\widehat{f}$ only
appears at quadratic and higher (even) orders in the dispersion relations. In \cite{Kostelecky:2013rta} it was stated that the corresponding component coefficients can be
removed from the physical observables by a field redefinition at first order in Lorentz violation, which reflects the results obtained here.
Nevertheless it is reasonable to investigate the properties of this operator, because it comprises the simplest set of
higher-dimensional {\em CPT}-odd component coefficients. An alternative would be the scalar operator $\widehat{e}$ in Tab.~I of \cite{Kostelecky:2013rta}.
However its properties are expected to be similar to the properties of $\widehat{m}$, which were already considered.

\subsubsection{Possible connection to the operator \texorpdfstring{$\widehat{c}^{\,\mu}$}{hatcmu}}
\label{sec:connection-cmu-f}

In \cite{Altschul:2006ts} it was shown that all Lorentz-violating modifications in observables depending on the minimal
coefficients $f^{(4)\alpha_1}$ cannot be distinguished from those of the minimal $c^{(4)\mu\alpha_1}$ coefficients. By a spinor
transformation the following exact correspondence was proven to be valid:
\begin{subequations}
\begin{equation}
\label{eq:identity-coefficients-c6}
c^{(4)\alpha_1\mu}=\frac{f^{(4)\alpha_1}f^{(4)\mu}}{(f^{(4)})^2}\left(\sqrt{1-(f^{(4)})^2}-1\right)\,,
\end{equation}
with the leading-order result
\begin{equation}
\label{eq:identity-coefficients-c6-expanded}
c^{\,(4)\alpha_1\mu}=-\frac{1}{2}f^{(4)\alpha_1}f^{(4)\mu}+\dots\,.
\end{equation}
\end{subequations}
Whether or not such an analogous equation is valid for the nonminimal $\widehat{c}^{\,(6)\mu}$ and $\widehat{f}^{(6)}$ could be
investigated by looking at the fermion energies for corresponding sets of coefficients of both operators. For example, the leading-order expansions of
Eqs.~(\ref{eq:dispersion-law-temporal-cmu-1}), (\ref{eq:dispersion-temporal-fmu-1}) are
\begin{equation}
\label{eq:connection-cmu-f-energy-1}
\widetilde{E}_{\psi,\widehat{c}^{\,\mu}}=E_{\psi}-c^{(6)0000}E_{\psi}^3\,,\quad \widetilde{E}_{\psi,\widehat{f}}=E_{\psi}+\frac{1}{2}(f^{(6)000})^2E_{\psi}^5\,,
\end{equation}
where similar expansions of Eqs.~(\ref{eq:dispersion-law-temporal-cmu-3}), (\ref{eq:dispersion-temporal-fmu-2}) result in
\begin{equation}
\widetilde{E}_{\psi,\widehat{c}^{\,\mu}}=E_{\psi}-c^{(6)ij00}p^ip^jE_{\psi}\,,\quad \widetilde{E}_{\psi,\widehat{f}}=E_{\psi}+\frac{1}{2}(f^{(6)i00}p^i)(f^{(6)j00}p^j)E_{\psi}^3\,.
\end{equation}
Finally, the leading-order expansions of Eqs.~(\ref{eq:dispersion-law-cmu-spatial}), (\ref{eq:dispersion-spatial-fmu}) read as follows:
\begin{equation}
\widetilde{E}_{\psi,\widehat{c}^{\,\mu}}=E_{\psi}-\frac{c^{(6)ijkl}p^ip^jp^kp^l}{E_{\psi}}\,,\quad \widetilde{E}_{\psi,\widehat{f}}=E_{\psi}+\frac{(f^{(6)ijk}p^ip^jp^k)(f^{(6)lmn}p^lp^mp^n)}{2E_{\psi}}\,.
\end{equation}
However, by comparing $\widetilde{E}_{\psi,\widehat{c}^{\,\mu}}$ and $\widetilde{E}_{\psi,\widehat{f}}$ it becomes evident that the energies
for each set of coefficients differ from each other by their energy-momentum dependence. For example, in \eqref{eq:connection-cmu-f-energy-1}
the Lorentz-violating modification in $\widetilde{E}_{\psi,\widehat{c}^{\,\mu}}$ depends on the third power of the fermion energy, whereas in
$\widetilde{E}_{\psi,\widehat{f}}$ it depends on the fifth power. Hence a relation analogous to Eqs. (\ref{eq:identity-coefficients-c6}),
(\ref{eq:identity-coefficients-c6-expanded}) cannot hold for the nonminimal operators $\widehat{c}^{\,(6)\mu}$ and $\widehat{f}^{(6)}$. There may
be the possibility of finding a connection between these operators by a field redefinition according to the lines of \cite{Kostelecky:2013rta},
which is not within the scope of the current article, though.

One consequence of such a correspondence (if it exists) would be that the {\em CPT}-odd handedness of the operator $\widehat{f}$ is
fictitious. Instead, the latter operator would have to be {\em CPT}-even because otherwise it would not be possible to transform it to
the {\em CPT}-even operator $\widehat{c}^{\,\mu}$. This apparent contradiction would be resolved when taking
into account that the operator mediating the {\em CPT} transformation may be changed in the presence of $\widehat{f}$ (cf. the remark in
parentheses above Eq. (23) in \cite{Altschul:2006ts}).

There is a further interesting fact on the minimal coefficients $f^{(4)\alpha_1}$ in the context of classical Lagrangians.
In \cite{Kostelecky:2010hs} the Lagrangians of a classical pointlike particle obeying the Lorentz-violating kinematics were obtained for
certain minimal coefficients. If all coefficients vanish except of the $f^{(4)\alpha_1}$, the Lagrangian of their Eq.~(8) only
depends on a quadratic combination of these coefficients. This is in accordance to the structure of the dispersion relations obtained
in the previous section where only quadratic powers of the nonminimal coefficients $f^{(6)\alpha_1\alpha_2\alpha_3}$ appear.

\subsubsection{Modified spinors and sums over spinor matrices}

For the particular case of the pseudoscalar operator $\widehat{f}$ the Dirac equation cannot only be block-diagonalized with the
matrix $U^{(1)}$ of \eqref{eq:diagonalization-matrix-first-order} but it can be diagonalized completely. Because of this the
positive-energy spinors at first order Lorentz violation can be obtained with the Hermitian conjugate of the matrix $U^{(1)}$.
They are given by
\begin{subequations}
\begin{align}
u^{(\alpha)}(\widetilde{E}_{\psi}^{(>)},\mathbf{p})&=\frac{1}{\sqrt{N_u^{(\alpha)}}}U^{(1)\,\dagger}(E_{\psi}^{(>)},m_{\psi},\mathbf{p})u^{(\alpha)}(m_{\psi},\mathbf{0})\,, \\[2ex]
\widehat{\mathcal{Q}}&=\widehat{\mathcal{Q}}(\widetilde{E}_{\psi}^{(>)},\mathbf{p})\,,\quad (p^{\mu})=(\widetilde{E}_{\psi}^{(>)},\mathbf{p})^T\,.
\end{align}
\end{subequations}
The negative-energy spinors read
\begin{subequations}
\begin{align}
v^{(\alpha)}(\widetilde{E}_{\psi}^{(>)},\mathbf{p})&=\frac{1}{\sqrt{N_v^{(\alpha)}}}U^{(1)\,\dagger}(E_{\psi}^{(>)},m_{\psi},-\mathbf{p})v^{(\alpha)}(m_{\psi},\mathbf{0})\,, \\[2ex]
\widehat{\mathcal{Q}}&=\widehat{\mathcal{Q}}(-\widetilde{E}_{\psi}^{(>)},-\mathbf{p})=-\widehat{\mathcal{Q}}(\widetilde{E}_{\psi}^{(>)},\mathbf{p})\,,\quad (p^{\mu})=(-\widetilde{E}_{\psi}^{(>)},-\mathbf{p})^T\,.
\end{align}
\end{subequations}
The spinor normalizations $N_u^{(\alpha)}$ and $N_v^{(\alpha)}$ are chosen analogously to \eqref{eq:spinor-normalization-conditions-cmu}.
The explicit expressions for the spinors are obtained in Sec.\ref{sec:spinors-completeness-explicit-fmu}. In the latter section of the
appendix the following sums over the positive- and negative-energy spinor matrices are deduced as well:
\begin{subequations}
\label{eq:spinor-completeness-fmu}
\begin{align}
\sum_{\alpha=1,2} u^{(\alpha)}(p)\overline{u}^{(\alpha)}(p)&=\cancel{p}+m_{\psi}\mathds{1}_4+\widehat{\mathcal{Q}}+\mathcal{O}(\widehat{f}^{\,2})\,, \\[2ex]
\sum_{\alpha=1,2} v^{(\alpha)}(p)\overline{v}^{(\alpha)}(p)&=\cancel{p}-m_{\psi}\mathds{1}_4-\widehat{\mathcal{Q}}+\mathcal{O}(\widehat{f}^{\,2})\,,
\end{align}
\end{subequations}
where $(p^{\mu})=(\widetilde{E}_{\psi}^{(>)},\mathbf{p})^T$ on the right-hand sides of the latter relations.
For the negative-energy spinors both $m_{\psi}\mathds{1}_4$ and $\widehat{\mathcal{Q}}$ come with a minus sign, which is crucial for
the validity of the optical theorem. The propagator is derived as usual by inverting the Dirac operator in momentum space,
$S^{-1}(p)=\cancel{p}-m_{\psi}\mathds{1}_4+\widehat{\mathcal{Q}}$, and expressing the result via the Dirac matrices needed. For the
case considered the \textit{Ansatz}
\begin{equation}
S(p)=a_1\gamma^0+a_2\gamma^1+a_3\gamma^2+a_4\gamma^3+a_5\mathds{1}_4+a_6\gamma^5\,,
\end{equation}
is sufficient because these are the Dirac matrices that appear in the Dirac operator. Solving the resulting linear system of equations with respect to the
variables $a_i$ leads to the modified propagator:
\begin{equation}
\label{eq:propagator-fhat}
S(p)=\frac{\mathrm{i}}{p^2-(m_{\psi}^2+\widehat{f}^{\,2})}\left(\cancel{p}+m_{\psi}\mathds{1}_4+\widehat{\mathcal{Q}}\right)=\frac{\mathrm{i}}{p^2-m_{\psi}^2}\left(\cancel{p}+m_{\psi}\mathds{1}_4+\widehat{\mathcal{Q}}\right)+\mathcal{O}(\widehat{f}^{\,2})\,.
\end{equation}
With the latter result and the sums over the spinor matrices of \eqref{eq:spinor-completeness-fmu} the validity of the optical theorem
at tree-level can be demonstrated for both electrons and positrons. The proof works such as for the case of the operator $\widehat{m}$,
cf. \secref{sec:fermion-propagator}. Note that with the expressions given the proof can only be done at first order Lorentz violation.
The spurious fermion dispersion relations, which may spoil the validity of the optical theorem for the temporal and mixed sector of the
coefficients considered, are removed at first order in Lorentz violation according to \secref{sec:dispersion-laws-fmu}. For the replacements of
Eqs.~(\ref{eq:dispersion-law-fmu-1-spurious-removed-1}), (\ref{eq:dispersion-law-fmu-1-spurious-removed-2}), and
(\ref{eq:dispersion-law-fmu-1-spurious-removed-3}) the proof works similarly.

The expansion of the propagator in $\widehat{f}^{\,2}$ after the second equality sign in \eqref{eq:propagator-fhat}
was performed to demonstrate the validity of the optical theorem at leading order in the component coefficients. However when computing amplitudes of
particle physics processes the full propagator should be taken into account. The reason is that the neglected contribution in the denominator is
important when the propagator is close to be on-shell.

\subsubsection{Analysis of unitarity}

\begin{figure}[t!]
\subfloat[]{\label{fig:plots-integrand-reflection-positivity-f-1}\includegraphics[scale=0.8]{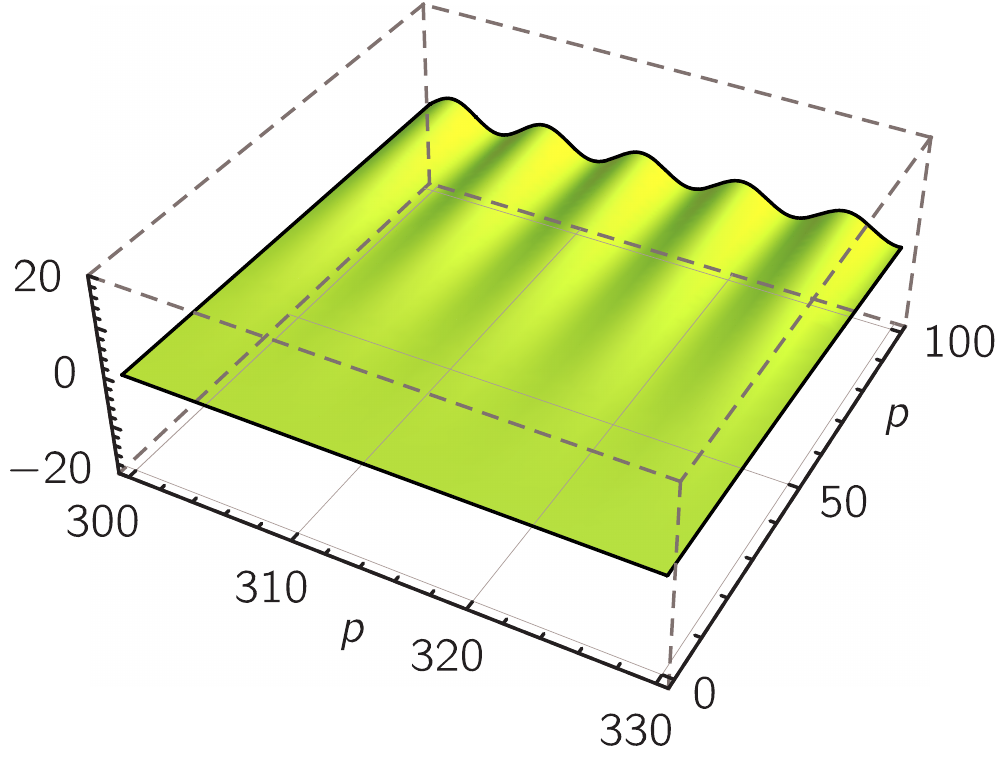}}
\subfloat[]{\label{fig:plots-integrand-reflection-positivity-f-2}\includegraphics[scale=0.8]{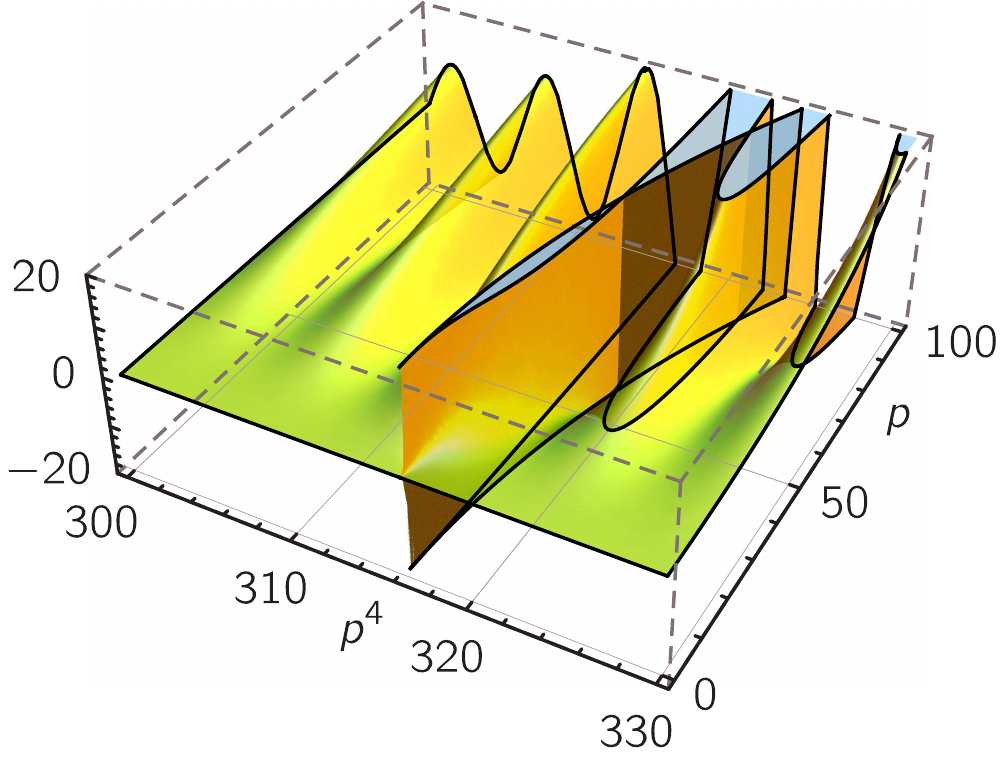}}
\caption{Plots of the integrand of \eqref{eq:reflection-positivity-temporal-f} as a function of $p^4$ and $p$ with the choices $m_{\psi}=1$, $x^4=1$
for distinct values of the Lorentz-violating component coefficient; (a) for $f^{(6)000}=0$ and (b) for $f^{(6)000}=10^{-5}$.}
\label{fig:plots-integrand-reflection-positivity-f}
\end{figure}

Finally reflection positivity and unitarity shall be investigated for the operator $\widehat{f}$. This will be done
according to the lines of Secs. \ref{sec:analysis-unitarity-mhat} and \ref{sec:analysis-unitarity-chatmu}. The considerations are restricted
to the single component coefficient $f^{(6)000}$ where the remaining Lorentz-violating coefficients are set to zero. First of all the Wick-rotated
scalar propagator part is given by:
\begin{equation}
S_E(p^4,\mathbf{p})=\frac{1}{(p^4)^2[1-(p^4)^4(f^{(6)000})^2]+\mathbf{p}^2+m_{\psi}^2}\,.
\end{equation}
Now what has to be computed is
\begin{equation}
\label{eq:reflection-positivity-temporal-f}
S_E(x^4,\mathbf{x})=8\pi \int_0^{\infty} \mathrm{d}p\,p^2 \int_0^{\infty} \mathrm{d}p^4\,\frac{\cos(p^4x^4)}{(p^4)^2[1-(p^4)^4(f^{(6)000})^2]+p^2+m_{\psi}^2}\,,
\end{equation}
just as in \eqref{eq:reflection-positivity-temporal-cmu} where spherical coordinates with $p\equiv |\mathbf{p}|$ are used. Again the denominator
of the Euclidean propagator involves
a polynomial of degree six in $p^4$, which makes it complicated to compute the integral analytically. Besides there appears a new issue that
did not occur for the particular component coefficients of the operators $\widehat{m}$ and $\widehat{c}^{\,\mu}$ considered. The form of the
denominator reveals that it can vanish for certain $p^4$, amongst them a positive value. Therefore for fixed $p$ the integrand has a pole
for $p^4\in [0,\infty)$. Such a pole appears for an arbitrarily small $f^{(6)000}$; it then resides at arbitrarily large $p^4$. Figures
\ref{fig:plots-integrand-reflection-positivity-f-1} and \ref{fig:plots-integrand-reflection-positivity-f-2} serve the purpose of illustrating
this behavior. In \figref{fig:plots-integrand-reflection-positivity-f-2} the poles are clearly visible as regions where the integrand diverges.

The question how to handle this pole remains an interesting open issue for future research. Since a standard Euclidean propagator
does not have any poles, the infinitesimal imaginary part used for Feynman's $\mathrm{i}\epsilon$-prescription is usually omitted in the
Euclidean propagator. Reinstating this infinitesimal imaginary part might lead to the definition of suitable integration contours with which
the poles can be avoided.

An alternative possibility for fixed $k$ would be to perform an integration in light of Cauchy's principal value. According to
this procedure a symmetric interval around the pole is cut out and after that the integration is performed. This method was applied numerically
for particular values of $x^4$, $m_{\psi}$, $f^{(6)000}$, and $p$. The numerical results for the integration over $p^4$ were always positive.
However since it is not clear whether this procedure is justifiable in this case, the numbers will not be stated. Furthermore the technique would
have to be extended to the full two-dimensional integration domain.

In \cite{Altschul:2006ts} it was stated that unitarity breaks down for the minimal coefficients, if $f^{(4)\mu}f^{(4)}_{\mu}>1$.
In case that the pole cannot be avoided, the behavior of the integrand may, indeed, be the nonminimal analogue of a violation of unitarity for the
operator $\widehat{f}$.

\section{Conclusion and outlook}
\label{sec:conclusion}
\setcounter{equation}{0}

To summarize, in the current article certain properties of quantum field theories that are based on the nonminimal spin-1/2 fermion sector of
the Lorentz-violating Standard-Model Extension were examined. For two {\em CPT}-even and one {\em CPT}-odd operator
(denoted as $\widehat{m}$, $\widehat{c}^{\,\mu}$, and $\widehat{f}$)
the modified fermion dispersion relations, the spinors, the sums over the matrices formed from the spinors, and the fermion propagator were obtained. For some subsets
of component coefficients spurious dispersion laws emerge that are not a perturbation of the standard dispersion relation. It was demonstrated
that these can be removed at first order in Lorentz violation. Furthermore the validity of the optical theorem at tree-level for both fermions and
antifermions was proven. For the {\em CPT}-even operators the proof is exact in the Lorentz-violating parameters, whereas for the
{\em CPT}-odd case it was performed at first order in Lorentz violation.

For some particular component coefficients unitarity of the modified quantum field theory was checked with the proporty of
reflection positivity. The result is that unitarity can be granted for the {\em CPT}-even operator $\widehat{m}$ considered.
For special component coefficients of the operator $\widehat{c}^{\,\mu}$ issues with unitarity arise. For the {\em CPT}-odd operator
$\widehat{f}$ it is not even clear how to apply the method of checking reflection positivity due to further problems. Substantiating
these results and working out a solution to the problems is beyond the scope of the paper and it remains an important task for future studies.

To conclude, in the framework of the analysis performed no issues were found for the operator $\widehat{m}$. Hence the latter
seems to result in a well-behaved quantum field theory. However this cannot be said about at least some of the component coefficients
of $\widehat{c}^{\,\mu}$ and $\widehat{f}$. The spinors, sums of the spinor matrices, and propagators determined for nonzero
$\widehat{m}$ can be used in upcoming particle physics calculations related to phenomenology. In contrast, the operators
$\widehat{c}^{\,\mu}$ and $\widehat{f}$ should be considered with care due to the issues arising in the context of unitarity. A further future
goal is to apply the methods demonstrated to investigate further operators that were not considered in this paper. The
fermion sector of the nonminimal SME especially, is still a {\em terra incognita} for both experiment \cite{Kostelecky:2013dka} and theory.

\section{Acknowledgments}

It is a pleasure to thank V.~A.~Kosteleck\'{y} for helpful discussions and suggestions. Furthermore the author is indebted
to both referees for their useful advice, which has improved the contents of the paper. Special thanks is due to John Hall for proofreading
parts of the article. This work was performed with financial support from the \textit{Deutsche Akademie der Naturforscher Leopoldina} within
Grant No. LPDS 2012-17.

\begin{appendix}
\numberwithin{equation}{section}

\section{Explicit spinors and sums of spinor matrices}
\setcounter{equation}{0}

In the following sections the explicit expressions for the modified Dirac spinors shall be obtained for the various sectors considered.
It is convenient to perform the calculations with the matrices and spinors in $2\times 2$ block form at first.

\subsection{\textit{CPT}-even scalar operator \texorpdfstring{$\boldsymbol{\widehat{m}}$}{hatm}}
\label{sec:spinors-completeness-explicit-mmu}

The explicit positive-energy spinors can be obtained directly from \eqref{eq:positive-energy-spinors} by using the Hermitian conjugate of
the transformation matrix $U$ given in \eqref{eq:diagonalization-matrix}. The latter reads
\begin{subequations}
\begin{equation}
U^{\dagger}=\begin{pmatrix}
\mathbf{n}\cdot\boldsymbol{\sigma}+n^0\mathds{1}_2 & \mathbf{n}\cdot\boldsymbol{\sigma}-n^0\mathds{1}_2 \\
-\mathbf{n}\cdot\boldsymbol{\sigma}+n^0\mathds{1}_2 & \mathbf{n}\cdot\boldsymbol{\sigma}+n^0\mathds{1}_2 \\
\end{pmatrix}\,,
\end{equation}
where $\boldsymbol{\sigma}=(\sigma^1,\sigma^2,\sigma^3)$ with the Pauli matrices $\sigma^1$, $\sigma^2$, and $\sigma^3$ of
\eqref{eq:pauli-matrices}. For convenience the four-vector $(n^{\mu})=(n^0,\mathbf{n})$ is introduced with the following components:
\begin{equation}
\label{eq:constants-mhat-case}
\mathbf{n} \equiv\begin{pmatrix}
\alpha \\
\beta \\
\gamma \\
\end{pmatrix}=-\frac{\mathbf{p}}{2\sqrt{\widetilde{E}_{\psi}(\widetilde{E}_{\psi}+\widetilde{m}_{\psi})}}\,,\quad n^0\equiv \delta=\frac{\sqrt{\widetilde{E}_{\psi}(\widetilde{E}_{\psi}+\widetilde{m}_{\psi})}}{2\widetilde{E}_{\psi}}\,.
\end{equation}
\end{subequations}
With these quantities the positive-energy spinors can be cast in the form
\begin{subequations}
\begin{align}
u^{(1)}(p)&=\frac{1}{\sqrt{N_u^{(1)}}}U^{\dagger}(\widetilde{E}^{(>)}_{\psi},\widetilde{m}_{\psi},\mathbf{p})\begin{pmatrix}
\phi^{(1)} \\
\mathbf{0} \\
\end{pmatrix}=\frac{1}{\sqrt{N_u^{(1)}}}\begin{pmatrix}
\gamma+\delta \\
\alpha+\mathrm{i}\beta \\
-\gamma+\delta \\
-\alpha-\mathrm{i}\beta \\
\end{pmatrix}\,, \\[2ex]
u^{(2)}(p)&=\frac{1}{\sqrt{N_u^{(2)}}}U^{\dagger}(\widetilde{E}^{(>)}_{\psi},\widetilde{m}_{\psi},\mathbf{p})\begin{pmatrix}
\phi^{(2)} \\
\mathbf{0} \\
\end{pmatrix}=\frac{1}{\sqrt{N_u^{(2)}}}\begin{pmatrix}
\alpha-\mathrm{i}\beta \\
-\gamma+\delta \\
-\alpha+\mathrm{i}\beta \\
\gamma+\delta \\
\end{pmatrix}\,, \\[2ex]
N_u^{(1)}&=N_u^{(2)}=\frac{1}{\widetilde{m}_{\psi}}n^2\,.
\end{align}
\end{subequations}
Now the sum over the positive-energy spinor matrices results in
\begin{subequations}
\begin{align}
\sum_{\alpha=1,2} u^{(\alpha)}(p)\overline{u}^{(\alpha)}(p)&=\frac{\widetilde{m}_{\psi}}{n^2}\begin{pmatrix}
M_0 & 0 & M_+ & -M^{*} \\
0 & M_0 & -M & M_- \\
M_- & M^{*} & M_0 & 0 \\
M & M_+ & 0 & M_0 \\
\end{pmatrix}\,, \\[2ex]
M_0&=n^2\,, \\[2ex]
\label{eq:matrix-elements}
M_+&=\alpha^2+\beta^2+(\gamma+\delta)^2\,,\quad M_-=\alpha^2+\beta^2+(\gamma-\delta)^2\,, \\[2ex]
\label{eq:matrix-elements-2}
M&=-2(\alpha+\mathrm{i}\beta)\delta\,,\quad M^{*}=-2(\alpha-\mathrm{i}\beta)\delta\,.
\end{align}
\end{subequations}
From the determinant condition it follows that $\widetilde{E}_{\psi}^2-\widetilde{m}_{\psi}^2=\mathbf{p}^2$, which can be used to obtain
the positive-energy relation of \eqref{eq:completeness-relation-positive}.

The negative-energy spinors follow from \eqref{eq:negative-energy-spinors} and they are given by:
\begin{subequations}
\begin{align}
v^{(1)}(p)&=\frac{1}{\sqrt{N_v^{(1)}}}U^{\dagger}(\widetilde{E}^{(>)}_{\psi},\widetilde{m}_{\psi},-\mathbf{p})\begin{pmatrix}
\mathbf{0} \\
\chi^{(1)} \\
\end{pmatrix}=-\frac{1}{\sqrt{N_v^{(1)}}}\begin{pmatrix}
\gamma+\delta \\
\alpha+\mathrm{i}\beta \\
\gamma-\delta \\
\alpha+\mathrm{i}\beta \\
\end{pmatrix}\,, \\[2ex]
v^{(2)}(p)&=\frac{1}{\sqrt{N_v^{(2)}}}U^{\dagger}(\widetilde{E}^{(>)}_{\psi},\widetilde{m}_{\psi},-\mathbf{p})\begin{pmatrix}
\mathbf{0} \\
\chi^{(2)} \\
\end{pmatrix}=-\frac{1}{\sqrt{N_v^{(2)}}}\begin{pmatrix}
\alpha-\mathrm{i}\beta \\
-\gamma+\delta \\
\alpha-\mathrm{i}\beta \\
-(\gamma+\delta) \\
\end{pmatrix}\,, \\[2ex]
N_v^{(1)}&=N_v^{(2)}=\frac{1}{\widetilde{m}_{\psi}}n^2\,.
\end{align}
\end{subequations}
The sum over the negative-energy spinor matrices is then
\begin{subequations}
\begin{align}
\sum_{\alpha=1,2} v^{(\alpha)}(p)\overline{v}^{(\alpha)}(p)&=\frac{\widetilde{m}_{\psi}}{n^2}\begin{pmatrix}
\overline{M}_0 & 0 & M_+ & -M^{*} \\
0 & \overline{M}_0 & -M & M_- \\
M_- & M^{*} & \overline{M}_0 & 0 \\
M & M_+ & 0 & \overline{M}_0 \\
\end{pmatrix}\,, \\[2ex]
\overline{M}_0&=-n^2\,,
\end{align}
\end{subequations}
where $M_+$, $M_-$, $M$, and $M^{*}$ are given by Eqs. (\ref{eq:matrix-elements}), (\ref{eq:matrix-elements-2}). With \eqref{eq:constants-mhat-case}
this leads to the result of \eqref{eq:spinor-completeness-negative}.

\subsection{{\em CPT}-even vector operator \texorpdfstring{$\boldsymbol{\widehat{c}^{\,\mu}}$}{hatc}}
\label{sec:spinors-completeness-explicit-cmu}

For this operator the computations of the previous section can be performed completely analogously with the
replacements $\widetilde{m}_{\psi}\mapsto m_{\psi}$ plus $p^{\mu}\mapsto p^{\mu}+\widehat{c}^{\,\mu}$ for both the positive-energy and
the negative-energy spinors (but the momentum components in $\widehat{c}^{\,\mu}$ itself remain untouched, of course). With this knowledge
the relations of \eqref{eq:spinor-completeness-cmu} can be computed. Here it is convenient to use
$(\widetilde{E}_{\psi}+\widehat{c}^{\,0})^2-m_{\psi}^2=(\mathbf{p}+\widehat{\mathbf{c}})^2$, which is obtained from
\eqref{eq:off-shell-dispersion-law-chat}.

\subsection{{\em CPT}-odd pseudoscalar operator \texorpdfstring{$\boldsymbol{\widehat{f}}$}{hatf}}
\label{sec:spinors-completeness-explicit-fmu}

In this case the diagonalization matrix $U$ is computed at first order in Lorentz violation. It results from \eqref{eq:diagonalization-matrix-first-order}
and its Hermitian conjugate is explicitly given by:
\begin{subequations}
\begin{align}
U^{(1)\,\dagger}&=\begin{pmatrix}
\mathbf{n}\cdot\boldsymbol{\sigma}+n^0\mathds{1}_2 & \mathbf{n}\cdot\boldsymbol{\sigma}-n^0\mathds{1}_2 \\
-\mathbf{n}^{*}\cdot\boldsymbol{\sigma}+(n^0)^{*}\mathds{1}_2 & \mathbf{n}^{*}\cdot\boldsymbol{\sigma}+(n^0)^{*}\mathds{1}_2 \\
\end{pmatrix} \displaybreak[0]\\[2ex]
\label{eq:constants-cpt-odd-case}
\mathbf{n}&\equiv\begin{pmatrix}
\alpha \\
\beta \\
\gamma \\
\end{pmatrix}=-\frac{\mathbf{p}\,\mathcal{C}}{8E_{\psi}^{5/2}(E_{\psi}+m_{\psi})^{3/2}}\,,\quad n^0\equiv \delta=\frac{\mathcal{C}^{*}}{8E_{\psi}^2\sqrt{E_{\psi}(E_{\psi}+m_{\psi})}}\,, \displaybreak[0]\\[2ex]
\mathcal{C}&=4E_{\psi}^3+2\mathrm{i}\widehat{f}^{\,(6)}E_{\psi}m_{\psi}+\mathrm{i}\widehat{f}^{\,(6)}(\mathbf{p}^2+m_{\psi}^2)+E_{\psi}^2(4m_{\psi}+\mathrm{i}\widehat{f}^{\,(6)})\,, \displaybreak[0]\\[2ex]
\widehat{f}^{\,(6)}&=f^{(6)\alpha_1\alpha_2\alpha_3}p_{\alpha_1}p_{\alpha_2}p_{\alpha_3}\,.
\end{align}
\end{subequations}
Note that the latter formulae involve the standard fermion energy $E_{\psi}$ instead of the modification $\widetilde{E}_{\psi}$.
The positive-energy spinors then read as
\begin{subequations}
\begin{align}
u^{(1)}(p)&=\frac{1}{\sqrt{N_u^{(1)}}}U^{(1)\,\dagger}(E^{(>)}_{\psi},m_{\psi},\mathbf{p})\begin{pmatrix}
\phi^{(1)} \\
\mathbf{0} \\
\end{pmatrix}=\frac{1}{\sqrt{N_u^{(1)}}}\begin{pmatrix}
\gamma+\delta \\
\alpha+\mathrm{i}\beta \\
-\gamma^{*}+\delta^{*} \\
-\alpha^{*}-\mathrm{i}\beta^{*} \\
\end{pmatrix}\,, \displaybreak[0]\\[2ex]
u^{(2)}(p)&=\frac{1}{\sqrt{N_u^{(2)}}}U^{(1)\,\dagger}(E^{(>)}_{\psi},m_{\psi},\mathbf{p})\begin{pmatrix}
\phi^{(2)} \\
\mathbf{0} \\
\end{pmatrix}=\frac{1}{\sqrt{N_u^{(2)}}}\begin{pmatrix}
\alpha-\mathrm{i}\beta \\
-\gamma+\delta \\
-\alpha^{*}+\mathrm{i}\beta^{*} \\
\gamma^{*}+\delta^{*} \\
\end{pmatrix}\,, \displaybreak[0]\\[2ex]
N_u^{(1)}&=N_u^{(2)}=\frac{1}{m_{\psi}}\left[(\mathrm{Re}\,n^0)^2-(\mathrm{Im}\,n^0)^2-(\mathrm{Re}\,\boldsymbol{n})^2+(\mathrm{Im}\,\boldsymbol{n})^2\right]\,,
\end{align}
\end{subequations}
and the sum over the positive-energy spinor matrices is given by:
\begin{subequations}
\begin{align}
\sum_{\alpha=1,2} u^{(\alpha)}(p)\overline{u}^{(\alpha)}(p)&=\frac{1}{N_u^{(1)}}\begin{pmatrix}
M_0 & 0 & M_+ & -M^{*} \\
0 & M_0 & -M & M_- \\
M_- & M^{*} & M_0^{*} & 0 \\
M & M_+ & 0 & M_0^{*} \\
\end{pmatrix}\,, \\[2ex]
\label{eq:matrix-elements-cpt-odd-case}
M_0&=n^2\,,\quad M_0^{*}=(n^{*})^2\,, \\[2ex]
M_+&=|\alpha|^2+|\beta|^2-2\mathrm{Im}(\alpha\beta^{*})+|\gamma|^2+|\delta|^2+2\mathrm{Re}(\gamma\delta^{*})\,, \\[2ex] M_-&=|\alpha|^2+|\beta|^2+2\mathrm{Im}(\alpha\beta^{*})+|\gamma|^2+|\delta|^2-2\mathrm{Re}(\gamma\delta^{*})\,, \\[2ex]
M&=(\alpha^{*}+\mathrm{i}\beta^{*})(\gamma-\delta)-(\alpha+\mathrm{i}\beta)(\gamma^{*}+\delta^{*})\,, \\[2ex] M^{*}&=(\alpha-\mathrm{i}\beta)(\gamma^{*}-\delta^{*})-(\alpha^{*}-\mathrm{i}\beta^{*})(\gamma+\delta)\,.
\end{align}
\end{subequations}
With the composite operator of \eqref{eq:constants-cpt-odd-case} one can show that
\begin{align}
\sum_{\alpha=1,2} u^{(\alpha)}(p)\overline{u}^{(\alpha)}(p)&=\begin{pmatrix}
m_{\psi}-\mathrm{i}\widehat{f}^{\,(6)} & 0 & E_{\psi}-p_3 & -(p_1-\mathrm{i}p_2) \\
0 & m_{\psi}-\mathrm{i}\widehat{f}^{\,(6)} & -(p_1+\mathrm{i}p_2) & E_{\psi}+p_3 \\
E_{\psi}+p_3 & p_1-\mathrm{i}p_2 & m_{\psi}+\mathrm{i}\widehat{f}^{\,(6)} & 0 \\
p_1+\mathrm{i}p_2 & E_{\psi}-p_3 & 0 & m_{\psi}+\mathrm{i}\widehat{f}^{\,(6)} \\
\end{pmatrix}+\mathcal{O}[(\widehat{f}^{\,(6)})^2] \notag \\
&=\cancel{p}+m_{\psi}\mathds{1}_4+\mathrm{i}\widehat{f}^{\,(6)}\gamma^5+\mathcal{O}[(\widehat{f}^{\,(6)})^2]=\cancel{p}+m_{\psi}\mathds{1}_4+\widehat{\mathcal{Q}}+\mathcal{O}[(\widehat{f}^{\,(6)})^2]\,.
\end{align}
Now the negative-energy spinors are
\begin{subequations}
\begin{align}
v^{(1)}(p)&=\frac{1}{\sqrt{N_v^{(1)}}}U^{(1)\,\dagger}(E^{(>)}_{\psi},m_{\psi},-\mathbf{p})\begin{pmatrix}
\mathbf{0} \\
\chi^{(1)} \\
\end{pmatrix}=-\frac{1}{\sqrt{N_v^{(1)}}}\begin{pmatrix}
\gamma+\delta \\
\alpha+\mathrm{i}\beta \\
\gamma^{*}-\delta^{*} \\
\alpha^{*}+\mathrm{i}\beta^{*} \\
\end{pmatrix}\,, \displaybreak[0]\\[2ex]
v^{(2)}(p)&=\frac{1}{\sqrt{N_v^{(2)}}}U^{(1)\,\dagger}(E^{(>)}_{\psi},m_{\psi},-\mathbf{p})\begin{pmatrix}
\mathbf{0} \\
\chi^{(2)} \\
\end{pmatrix}=-\frac{1}{\sqrt{N_v^{(2)}}}\begin{pmatrix}
\alpha-\mathrm{i}\beta \\
-\gamma+\delta \\
\alpha^{*}-\mathrm{i}\beta^{*} \\
-(\gamma^{*}+\delta^{*}) \\
\end{pmatrix}\,, \displaybreak[0]\\[2ex]
N_v^{(1)}&=N_v^{(2)}=\frac{1}{m_{\psi}}\left[(\mathrm{Re}\,n^0)^2-(\mathrm{Im}\,n^0)^2-(\mathrm{Re}\,\boldsymbol{n})^2+(\mathrm{Im}\,\boldsymbol{n})^2\right]\,,
\end{align}
\end{subequations}
and with the matrix elements of \eqref{eq:matrix-elements-cpt-odd-case} and the results
of \eqref{eq:constants-cpt-odd-case} one obtains:
\begin{subequations}
\begin{align}
\hspace{-0.3cm}\sum_{\alpha=1,2} v^{(\alpha)}(p)\overline{v}^{(\alpha)}(p)&=\frac{1}{N_v^{(1)}}\begin{pmatrix}
-M_0 & 0 & M_+ & -M^{*} \\
0 & -M_0 & -M & M_- \\
M_- & M^{*} & -M_0^{*} & 0 \\
M & M_+ & 0 & -M_0^{*} \\
\end{pmatrix} \notag \displaybreak[0]\\
&=\begin{pmatrix}
-m_{\psi}+\mathrm{i}\widehat{f}^{\,(6)} & 0 & E_{\psi}-p_3 & -(p_1-\mathrm{i}p_2) \\
0 & -m_{\psi}+\mathrm{i}\widehat{f}^{\,(6)} & -(p_1+\mathrm{i}p_2) & E_{\psi}+p_3 \\
E_{\psi}+p_3 & p_1-\mathrm{i}p_2 & -m_{\psi}-\mathrm{i}\widehat{f}^{\,(6)} & 0 \\
p_1+\mathrm{i}p_2 & E_{\psi}-p_3 & 0 & -m_{\psi}-\mathrm{i}\widehat{f}^{\,(6)} \\
\end{pmatrix}+\mathcal{O}[(\widehat{f}^{\,(6)})^2] \notag \displaybreak[0]\\
&=\cancel{p}-m_{\psi}\mathds{1}_4-\mathrm{i}\widehat{f}^{\,(6)}\gamma^5+\mathcal{O}[(\widehat{f}^{\,(6)})^2]=\cancel{p}-m_{\psi}\mathds{1}_4-\widehat{\mathcal{Q}}+\mathcal{O}[(\widehat{f}^{\,(6)})^2]\,.
\end{align}
\end{subequations}
This completes the derivation of the results given by \eqref{eq:spinor-completeness-fmu}.

\end{appendix}

\newpage



\begin{thebibliography}{99}

\bibitem{Kostelecky:1988zi}
V.~A.~Kosteleck\'{y} and S.~Samuel,
``Spontaneous breaking of Lorentz symmetry in string theory,''
Phys.\ Rev.\ D {\bf 39}, 683 (1989).

\bibitem{Kostelecky:1991ak}
V.~A.~Kosteleck\'{y} and R.~Potting,
``\textit{CPT} and strings,''
Nucl.\ Phys.\ B {\bf 359}, 545 (1991).

\bibitem{Kostelecky:1994rn}
V.~A.~Kosteleck\'{y} and R.~Potting,
``\textit{CPT}, strings, and meson factories,''
Phys.\ Rev.\ D {\bf 51}, 3923 (1995),
hep-ph/9501341.

\bibitem{oai:arXiv.org:hep-th/9605088}
V.~A.~Kosteleck\'{y} and R.~Potting,
``Expectation values, Lorentz invariance, and \textit{CPT} in the open bosonic string,''
Phys.\ Lett.\ B {\bf 381}, 89 (1996),
hep-th/9605088.

\bibitem{Bojowald:2004bb}
M.~Bojowald, H.~A.~Morales-T\'{e}cotl, and H.~Sahlmann,
``On loop quantum gravity phenomenology and the issue of Lorentz invariance,''
Phys.\ Rev.\ D {\bf 71}, 084012 (2005),
gr-qc/0411101.

\bibitem{Gambini:1998it}
R.~Gambini and J.~Pullin,
``Nonstandard optics from quantum space-time,''
Phys.\ Rev.\ D {\bf 59}, 124021 (1999),
gr-qc/9809038.

\bibitem{Carroll:2001ws}
S.~M.~Carroll, J.~A.~Harvey, V.~A.~Kosteleck\'{y}, C.~D.~Lane, and T.~Okamoto,
``Noncommutative field theory and Lorentz violation,''
Phys.\ Rev.\ Lett.\ {\bf 87}, 141601 (2001),
hep-th/0105082.

\bibitem{Klinkhamer:2003ec}
F.~R.~Klinkhamer and C.~Rupp,
``Spacetime foam, \textit{CPT} anomaly, and photon propagation,''
Phys.\ Rev.\ D {\bf 70}, 045020 (2004),
hep-th/0312032.

\bibitem{Bernadotte:2006ya}
S.~Bernadotte and F.~R.~Klinkhamer,
``Bounds on length-scales of classical spacetime foam models,''
Phys.\ Rev.\ D {\bf 75}, 024028 (2007),
hep-ph/0610216.

\bibitem{Klinkhamer:1998fa}
F.~R.~Klinkhamer,
``Z-string global gauge anomaly and Lorentz non-invariance,''
Nucl.\ Phys.\ B {\bf 535}, 233 (1998),
hep-th/9805095.

\bibitem{Klinkhamer:1999zh}
F.~R.~Klinkhamer,
``A \textit{CPT} anomaly,''
Nucl.\ Phys.\ B {\bf 578}, 277 (2000),
hep-th/9912169.

\bibitem{Colladay:1998fq}
D.~Colladay and V.~A.~Kosteleck\'{y},
``Lorentz violating extension of the standard model,''
Phys.\ Rev.\ D {\bf 58}, 116002 (1998),
hep-ph/9809521.

\bibitem{Kostelecky:2008ts}
V.~A.~Kosteleck\'{y} and N.~Russell,
``Data Tables for Lorentz and \textit{CPT} Violation,''
Rev.\ Mod.\ Phys.\  {\bf 83}, 11 (2011),
arXiv:0801.0287 [hep-ph].

\bibitem{Kostelecky:2009zp}
V.~A.~Kosteleck\'{y} and M.~Mewes,
``Electrodynamics with Lorentz-violating operators of arbitrary dimension,''
Phys.\ Rev.\ D {\bf 80}, 015020 (2009),
arXiv:0905.0031 [hep-ph].

\bibitem{Kostelecky:2013rta}
V.~A.~Kosteleck\'{y} and M.~Mewes,
``Fermions with Lorentz-violating operators of arbitrary dimension,''
Phys.\ Rev.\ D {\bf 88}, 096006 (2013),
arXiv:1308.4973 [hep-ph].

\bibitem{Kostelecky:2011gq}
V.~A.~Kosteleck\'{y} and M.~Mewes,
``Neutrinos with Lorentz-violating operators of arbitrary dimension,''
Phys.\ Rev.\ D {\bf 85}, 096005 (2012),
arXiv:1112.6395 [hep-ph].

\bibitem{oai:arXiv.org:hep-ph/0101087}
C.~Adam and F.~R.~Klinkhamer,
``Causality and \textit{CPT} violation from an Abelian Chern-Simons-like term,''
Nucl.\ Phys.\ B {\bf 607}, 247 (2001),
hep-ph/0101087.

\bibitem{Casana-etal2009}
R.~Casana, M.~M.~Ferreira, A.~R.~Gomes, and P.~R.~D.~Pinheiro,
``Gauge propagator and physical consistency of the \textit{CPT}-even part of the
  standard model extension,''
Phys. Rev. D {\bf 80}, 125040 (2009),
arXiv:0909.0544 [hep-th].

\bibitem{Casana-etal2010}
R.~Casana, M.~M.~Ferreira, A.~R.~Gomes, and F.~E.~P.~dos Santos,
``Feynman propagator for the nonbirefringent \textit{CPT}-even
electrodynamics of the standard model extension,''
Phys.\ Rev.\ D {\bf 82}, 125006 (2010),
arXiv:1010.2776 [hep-th].

\bibitem{Klinkhamer:2010zs}
F.~R.~Klinkhamer and M.~Schreck,
``Consistency of isotropic modified Maxwell theory: Microcausality and unitarity,''
Nucl.\ Phys.\ B {\bf 848}, 90 (2011),
arXiv:1011.4258 [hep-th],
F.~R.~Klinkhamer and M.~Schreck,
``Models for low-energy Lorentz violation in the photon sector: Addendum to `Consistency of isotropic modified Maxwell theory',''
Nucl.\ Phys.\ B {\bf 856}, 666 (2012),
arXiv:1110.4101 [hep-th].

\bibitem{Schreck:2011ai}
M.~Schreck,
``Analysis of the consistency of parity-odd nonbirefringent modified Maxwell theory,''
Phys.\ Rev.\ D {\bf 86}, 065038 (2012),
arXiv:1111.4182 [hep-th].

\bibitem{Schreck:2013gma}
M.~Schreck,
``Quantum field theory based on birefringent modified Maxwell theory,''
Phys.\ Rev.\ D {\bf 89}, 085013 (2014),
arXiv:1311.0032 [hep-th].

\bibitem{Colladay:2014dua}
D.~Colladay, P.~McDonald, and R.~Potting,
``Gupta-Bleuler photon quantization in the SME,''
Phys.\ Rev.\ D {\bf 89}, 085014 (2014),
arXiv:1401.1173 [hep-ph].

\bibitem{Schreck:2013kja}
M.~Schreck,
``Quantum field theoretic properties of Lorentz-violating operators of nonrenormalizable dimension in the photon sector,''
Phys.\ Rev.\ D {\bf 89}, 105019 (2014),
arXiv:1312.4916 [hep-th].

\bibitem{Dirac:1928hu}
P.~A.~M.~Dirac,
``The Quantum theory of the electron,''
Proc.\ Roy.\ Soc.\ Lond.\ A {\bf 117}, 610 (1928).

\bibitem{Kostelecky:2000mm}
V.~A.~Kosteleck\'{y} and R.~Lehnert,
``Stability, causality, and Lorentz and {\em CPT} violation,''
Phys.\ Rev.\ D {\bf 63}, 065008 (2001),
hep-th/0012060.

\bibitem{Kostelecky:2007kx}
V.~A.~Kosteleck\'{y}, N.~Russell, and J.~D.~Tasson,
``Constraints on torsion from bounds on Lorentz violation,''
Phys.\ Rev.\ Lett.\ {\bf 100}, 111102 (2008),
arXiv:0712.4393 [gr-qc].

\bibitem{Kostelecky:2008in}
V.~A.~Kosteleck\'{y} and J.~D.~Tasson,
``Prospects for large relativity violations in matter-gravity couplings,''
Phys.\ Rev.\ Lett.\ {\bf 102}, 010402 (2009),
arXiv:0810.1459 [gr-qc].

\bibitem{Kostelecky:2010ze}
V.~A.~Kosteleck\'{y} and J.~D.~Tasson,
``Matter-gravity couplings and Lorentz violation,''
Phys.\ Rev.\ D {\bf 83}, 016013 (2011),
arXiv:1006.4106 [gr-qc].

\bibitem{Tasson:2012nx}
J.~D.~Tasson,
``Lorentz violation, gravitomagnetism, and intrinsic spin,''
Phys.\ Rev.\ D {\bf 86}, 124021 (2012),
arXiv:1211.4850 [hep-ph].

\bibitem{Tasson:2010nr}
J.~D.~Tasson,
``Gravitational physics with antimatter,''
Hyperfine Interact.\ {\bf 193}, 291 (2009),
arXiv:1010.2811 [hep-ph].

\bibitem{Tasson:2012au}
J.~D.~Tasson,
``Antimatter, the SME, and gravity,''
Hyperfine Interact.\ {\bf 213}, 137 (2012),
arXiv:1212.1636 [hep-ph].

\bibitem{Foldy:1949wa}
L.~L.~Foldy and S.~A.~Wouthuysen,
``On the Dirac theory of spin 1/2 particles and its non-relativistic limit,''
Phys.\ Rev.\  {\bf 78}, 29 (1950).

\bibitem{Cutkosky:1960sp}
R.~E.~Cutkosky,
``Singularities and discontinuities of Feynman amplitudes,''
J.\ Math.\ Phys.\  {\bf 1}, 429 (1960).

\bibitem{Mandl:1986}
F.~Mandl and G.~Shaw,
\textit{Quantum Field Theory}
(John Wiley \& Sons, Chichester $\cdot$ New York $\cdot$ Brisbane $\cdot$ Toronto $\cdot$ Singapore, 1986).

\bibitem{Montvay:1994}
I. Montvay und G. M\"{u}nster,
\textit{Quantum Fields on a Lattice},
(U.P., Cambridge, 1994).

\bibitem{Gradshteyn:2007}
I. S. Gradshteyn and I. M. Ryzhik,
\textit{Table of Integrals, Series, and Products}, 7nd ed.
(Academic Press, Burlington $\cdot$ San Diego $\cdot$ London, 2007).

\bibitem{Colladay:2001wk}
D.~Colladay and V.~A.~Kosteleck\'{y},
``Cross-sections and Lorentz violation,''
Phys.\ Lett.\ B {\bf 511}, 209 (2001),
hep-ph/0104300.

\bibitem{Altschul:2006ts}
B.~Altschul,
``Eliminating the {\em CPT}-odd $f$ coefficient from the Lorentz-violating standard model extension,''
J.\ Phys.\ A {\bf 39}, 13757 (2006),
hep-th/0602235.

\bibitem{Kostelecky:2010hs}
V.~A.~Kosteleck\'{y} and N.~Russell,
``Classical kinematics for Lorentz violation,''
Phys.\ Lett.\ B {\bf 693}, 443 (2010),
arXiv:1008.5062 [hep-ph].

\bibitem{Kostelecky:2013dka}
V.~A.~Kosteleck\'{y},
``Comments on Lorentz and {\em CPT} Violation,''
in V.~A.~Kosteleck\'{y}, ed., \textit{CPT and Lorentz Symmetry VI}
(World Scientific, Singapore, 2014),
arXiv:1309.3761 [hep-ph].

\end{thebibliography}
\end{document}